\documentclass[aps,onecolumn,floatfix,byrevtex,titlepage,balancelastpage,raggedbottom,amsfonts,amssymb,amsmath]{revtex4}
\usepackage{natbib}
\usepackage{epsfig}
\usepackage{bm}
\usepackage{times}
\usepackage{mathptmx}
\usepackage{amsmath}
\usepackage{graphicx}
\usepackage{graphics}
\usepackage{hyperref}
\usepackage{amssymb}
\usepackage{enumerate}
\usepackage{bm}
\usepackage[english]{babel}
\begin{document}
\bibliographystyle{plain}
\title{Introduction to Perl module lb2d.pm 0.1 for basic Lattice Boltzmann simulations}
\author{Danilo Sergi}
\affiliation{University of Applied Sciences (SUPSI),
The iCIMSI Research Institute,
Galleria 2, CH-6928 Manno, Switzerland}
\date{\today}
\begin{abstract}
The Lattice Boltzmann method has been gaining acceptance for hydrodynamic problems involving
complex boundaries, non-equilibrium dynamics and interfacial phenomena in multiphase or
multicomponent systems. lb2d.pm is a Perl module offering a clear framework for simple 
Lattice Boltzmann simulations in 2D. The syntax and some basic applications are presented. 
Mainly, the new version 0.1 enables solute transport and accounts for reactive boundaries
leading to surface growth.
\end{abstract}
\maketitle
\section{Introduction}

The present document aims at making viable the use of the module lb2d.pm \cite{lb2d}, version 0.1. This 
is a module for the scripting language Perl that allows to investigate readily simple systems by the 
Lattice Boltzmann method.

In the next section, we provide a brief account of the theory, along with the basic operations relevant
for computations.

In Sec.~\ref{sec:command} are listed all the commands, accompanied with a description. The commands are 
presented in the order they are more likely to be used in a general program. They can be organized into groups 
as follows:

Initialization: lattice, processors, read\_data, fluids, solute, inlet, outlet, inlet\_momentum,
outlet\_momentum, inlet\_solute, outlet\_solute, boundary\_style, obstacle, 
surface\_reaction, position0, momentum0, concentration0

Force fields: aveforce, aveacceleration, interforce, adhesiveforce, intersolute

Output: thermo, log, output, write\_restart

Action: iteration

This section should be read at least once completely from the beginning to the
end. The examples are given in Perl syntax. In the doubt that it could be
expected from us, we say that we tried to reproduce some aspects of the syntax of
the molecular dynamics code LAMMPS \cite{lmp}. The version 0.1 can treat solute transport and
surface reaction. The new commands are solute, inlet\_solute/outlet\_solute, surface\_reaction, concentration0 and
intersolute. The existing commands read\_data, fluids and output have been modified.

In the section Applications, we present some examples
worked out in the framework of the module lb2d.pm. In the section
Scripts, we discuss the programs used to generate and analyze the data
for the proposed case studies.

For more details on the Lattice Boltzmann method, the interested reader
is especially addressed to the books \cite{book1,book2}
\begin{enumerate}
\item[-]M.C.~Sukop and D.T.~Thorne Jr., Lattice Boltzmann Modelig: An
Introduction for Geoscientists and Engineers, Springer Verlang, Berlin
Heidelberg 2010.

\item[-]S.~Succi, The Lattice Boltzmann Equation for Fluid Dynamics and
Beyond, Oxford University Press, Oxford 2009.
\end{enumerate}

This work was committed by iCIMSI-SUPSI for research projects in the fields of mechanical 
engineering and materials science.

The module is freely available at \url{https://sites.google.com/site/lbmodule/}.
Questions, comments or suggestions can be sent to \url{lb.module@gmail.com}.

\clearpage
\section{\label{sec:model}Model and algorithm}

The Lattice Boltzmann method is based on the Lattice Boltzmann equation
\begin{equation}
f_{i}(\bm{r}+\bm{e}_{i}\Delta t,t+\Delta t)=\underbrace{f_{i}(\bm{r},t)}_{\mathrm{streaming}}
-\underbrace{\frac{1}{\tau}\big[f_{i}(\bm{r},t)-f_{i}^{\mathrm{eq}}(\bm{r},t)\big]}_{\mathrm{collision}}\ .
\label{eq:lb}
\end{equation}
$\tau$ is the relaxation time; $\bm{r}=(x,y)$ are the coordinates of the
node of a square lattice. The vector $\bm{e}_{i}$ defines the direction of
the $i$-th velocity mode. The distribution function of a velocity mode in
space and time is given by $f_{i}(\bm{r},t)$. In the above equation, the
first term on the right-hand side accounts for streaming, that is, free propagation, while the
second one describes collisions in the BGK approximation \cite{bgk}.

The number of velocity modes and the form of equilibrium distribution
functions depend on the specific model. In the d2q9 model, the systems are 2D
and there are nine velocity modes. In the module, the equilibrium distribution
functions are computed by means of the formula
\begin{equation}
f_{i}^{\mathrm{eq}}(\bm{r})=w_{i}\rho(\bm{r})\Big[1+\frac{\bm{e}_{i}\cdot\bm{u}}{c_{\mathrm{s}}^2}
+\frac{1}{2}\frac{(\bm{e}_{i}\cdot\bm{u})^{2}}{c_{\mathrm{s}}^{4}}-\frac{1}{2}\frac{\bm{u}^{2}}{c_{\mathrm{s}}^{2}}\Big]
\quad\quad i=0,\dots,8\ .
\label{eq:eq}
\end{equation}
The $w_{i}$'s are weighting factors given by $w_{0}=4/9$, $w_{i}=1/9$ for $i=1,2,3,4$ and $w_{i}=1/36$
for $i=5,6,7,8$. $c_{\mathrm{s}}=1/\sqrt{3}$ [lu/ts] is the speed of sound.
The density and velocity at a given node read respectively
\begin{equation*}
\rho(\bm{r})=\sum_{i=0}^{8}f_{i}(\bm{r})\quad\text{and}\quad
\bm{u}(\bm{r})=\frac{1}{\rho(\bm{r})}\sum_{i=0}^{8}f_{i}(\bm{r})\bm{e}_{i}\ .
\end{equation*}
To keep the notation simple, we omitted the time dependence.
\begin{figure}[b]
\includegraphics[height=7cm]{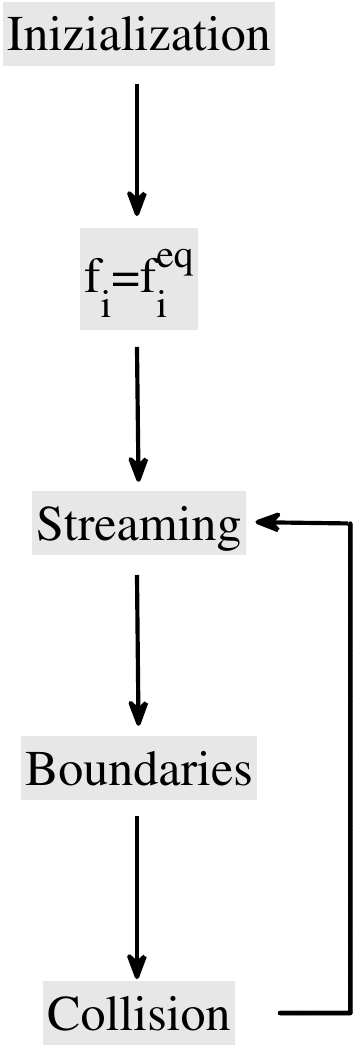}
\caption{
Steps of the algorithm for the Lattice Boltzmann dynamics as implemented in the module.
\label{fig:algorithm}}
\end{figure}

Figure \ref{fig:algorithm} outlines the basic steps of the method employed
for computations.
After initialization of the system, to every lattice site are attributed its
equilibrium distribution functions obtained from Eq.~\ref{eq:eq}.
Then, for every lattice site, free propagation is executed. If a velocity mode
reaches an obstacle, it is again propagated according to the selected boundary
style. (These two steps are performed inside the same loops and shall be referred
to as streaming step in the sequel.)
Finally, the equilibrium distribution functions are computed with
Eq.~\ref{eq:eq} and the collision term is evaluated. This scheme is repeated
from the streaming procedure for the number of timesteps.

\clearpage
\section{\label{sec:command}Commands}

\textbf{\underline{lattice command}}

\underline{Syntax:}

lattice(style,$N_{x}$,$N_{y}$,tau)
\begin{itemize}
\item style=d2q9
\item $N_{x}$,$N_{y}$=width of the simulation domain in the directions of $x$
  and $y$ axes
\item tau=relaxation time
\end{itemize}

\underline{Example:}

\&lattice(``d2q9'',100,100,1);

\underline{Description:}

This command defines the style of the underlying lattice and the size of the
rectangular simulation domain. The last argument fixes the relaxation time $\tau$.

\begin{figure}[b]
\includegraphics[height=5cm]{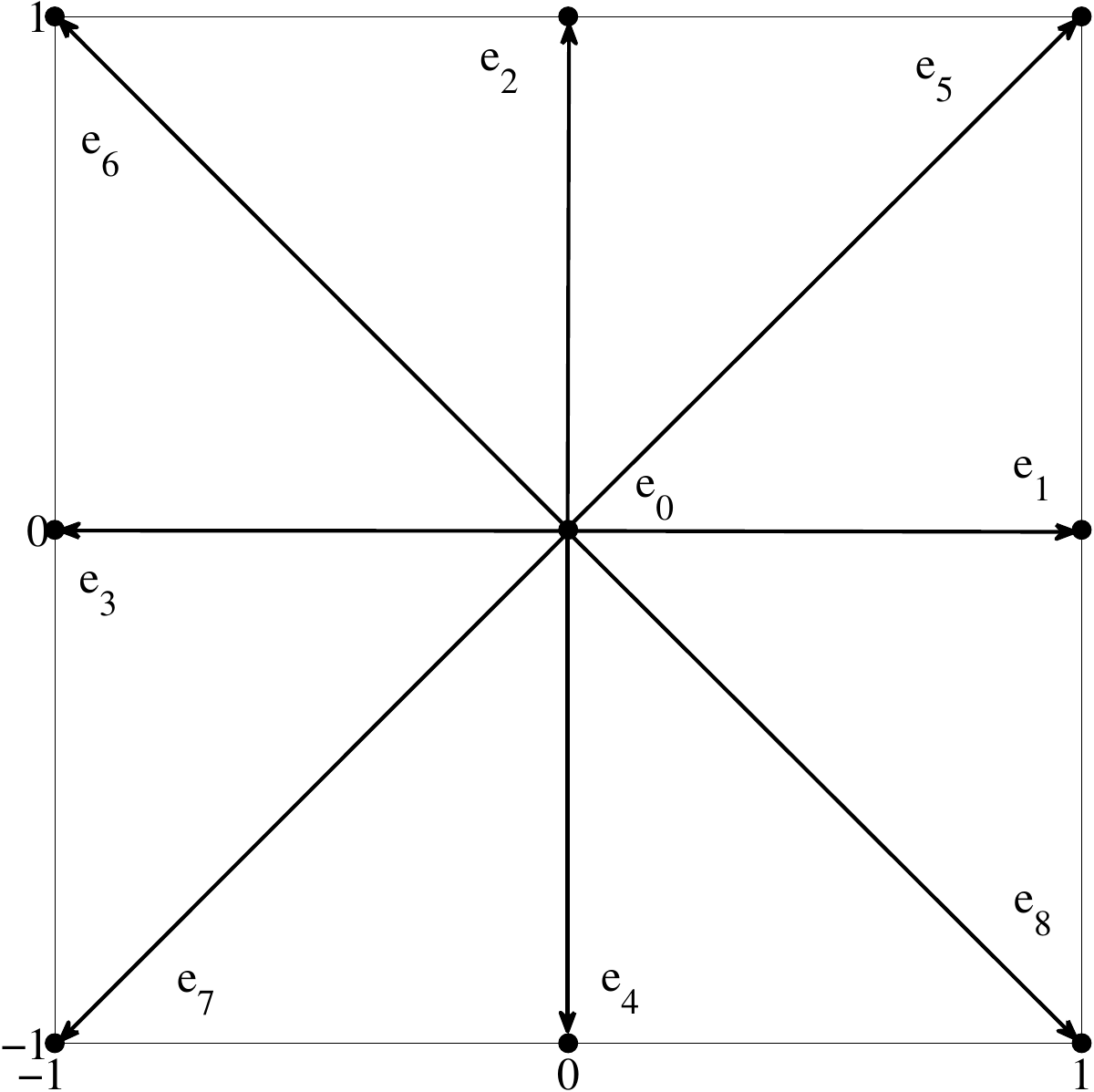}
\caption{
Velocity modes for the lattice d2q9.
\label{fig:lattice_d2q9}}
\end{figure}

For the style d2q9, the velocity modes are identified as shown in Fig.~\ref{fig:lattice_d2q9}.

The lattice sites are numbered from $0$ to $N_{x}$ ($N_{y}$).

By default, the boundaries of the simulation domain are periodic.

This command must appear in every script. If this command is used before
the read\_data command, the definitions of the last three arguments
are overridden if the sections relaxation\_time and domain are present in
the input data file. This command must precede the command processors,
otherwise the simulation is interrupted with an error message.

\underline{Related commands:}

processors read\_data
\vspace{0.5cm}

\textbf{\underline{processors command}}

\underline{Syntax:}

processors(px,py)
\begin{itemize}
\item px,py=$\sharp$ of processors along each axis
\end{itemize}

\underline{Example:}

\&processors(3,1);

\underline{Description:}

With this command the simulations are run in parallel. Specifically, the
initialization, streaming and collision steps of the basic algorithm
(see Fig.~\ref{fig:algorithm}) are executed using $p_{x}\times p_{y}$
processors. The principle of space decomposition of the simulation
domain is applied. For example, let $q_{x}$ and $r_{x}$ be integer numbers
such that $N_{x}=q_{x}p_{x}+r_{x}$. So, the $x$ side of every subdomain counts $q_{x}$
nodes with the exception of the last ones with $x\geq q_{x}(p_{x}-1)$:
in this case the side length is of $q_{x}+r_{x}+1$ nodes. The same rule
is applied for the other directions.

If this command is used with the serial version of the module, it
is disregarded. The parallel version of the module needs a multicore processor
and the ForkManager module \cite{fork}. This command must be used in every script
being executed with a parallel version of the module. This command must
precede any other command initializing density and momentum: i.e.,
inlet\_momentum/outlet\_momentum and momentum0. This command must follow the
lattice command.

Warnings: At line 2 of the parallel version of the module, the path of
the ForkManager module must be specified. On Windows operating systems
we incurred in registry problems.

\underline{Related command:}

lattice
\vspace{0.5cm}

\textbf{\underline{read\_data command}}

\underline{Syntax:}

read\_data(file,arguments)
\begin{itemize}
\item file=name of data file to read in
\item arguments=fluids,solute,equilibrium
\end{itemize}

\underline{Examples:}

\&read\_data(``../Input/data'');\\
\&read\_data(``restart.separation'',''fluids'');\\
\&read\_data(``restart.diffusion'',''solute'',''fluids'');\\
\&read\_data(``data.separation'',''equilibrium'',''fluids'');\\
\&read\_data(``data.diffusion'',''solute'',''equilibrium'',''fluids'');

\underline{Description:}

With this command it is defined the starting configuration of the system, partly or completely. The additional
arguments fluids, solute and equilibrium can be given in any order. With the additional argument equilibrium,
all distribution functions are replaced with their equilibrium values for both fluid flow and solute transport
(see Fig.~\ref{fig:algorithm}). With an additional argument of type fluids, the files defining the state of a 
two-component system are read in (see write\_restart command). For fluid flow, the information of the input data 
file must be organized into nine sections: relaxation\_time, domain, processors, inlet, outlet, boundary, field, density, momentum.

The relaxation\_time section is defined as follows:\\
relaxation\_time\\
$\tau$ $\alpha_{\sigma}$\\
In the case of multicomponent systems, it is possible to specify an additional parameter $\alpha_{\sigma}$ controlling
the density ratio between the substances $\sigma=1,2$ (see fluids command).

The domain section is defined as follows:\\
domain\\
$N_{x}$ $N_{y}$

The processors section is defined as follows:\\
processors\\
$p_{x}$ $p_{y}$\\
$p_{x}$ and $p_{y}$ define the grid of processors.

The inlet and/or outlet sections are defined as follows:\\
inlet\\
wall $x_{0}$ $x_{1}$ label style arguments\\
outlet\\
wall $x_{0}$ $x_{1}$ label style arguments\\
Details can be found in the explanation of the commands
inlet/outlet and inlet\_momentum/outlet\_momentum.

The boundary section is defined as follows:\\
boundary\\
style1 argument style2\\
noslip, slip, mixed and periodic are the available styles. An additional argument
is necessary for the mixed style with only one reflection coefficient.

The field section is defined as follows:\\
field\\
force mode1 value1 mode2 value2 mode3 value3 $\dots$\\
acceleration mode1 value1 mode2 value2 mode3 value3 $\dots$\\
interforce $G$ $\psi_{0}$ $\rho_{0}$\\
adhesiveforce $G$ $\psi_{0}$ $\rho_{0}$\\
mode's are the directions of the velocity modes, namely e1,$\dots$, e8;
value's are the associated intensities. The parameters for interforce
and adhesiveforce are explained in the description of the corresponding commands.

The density section is defined as follows:\\
density\\
$x$ $y$ $\rho$ boundary\_style \\
$x$ and $y$ are the coordinates of the lattice site and $\rho$ its density. boundary\_style
is one of the available styles for the treatment of collisions in the hybrid case (see boundary section as well as
boundary\_style and obstacle commands) because more than one style is used.

The momentum section is defined as follows:\\
momentum\\
$x$ $y$ $f_0$ $f_1$ $f_2$ $f_3$ $f_4$ $f_5$ $f_6$ $f_7$ $f_8$\\
$x$ and $y$ are the coordinates of the lattice site. For the d2q9 lattice
style, $f_0$,$\dots$, $f_8$ are the populations of the velocity modes.

A site is identified as solid boundary if its density is equal to $-11$.
The components of the momentum can take any value in this case. For clarity,
we suggest to set them to $0$.

The section domain must precede the section processors.
If present, the section field must follow the section relaxation\_time; the section
processors must precede the sections density and momentum; and the section boundary
must precede the section density.
In all other cases, the sections and their items can be given in any order.

If the read\_data command appears before the lattice command and the input file
contains the sections relaxation\_time and/or domain, this information is
overridden. If the input file contains the section processors, the command processors
does not redefine the grid of processors. It is possible to redefine the grid of processors
of a previous simulation by modifying manually the restart file.
If the input file contains force and/or acceleration in the section field,
the commands aveforce and aveacceleration in the script add further contributions to the applied forces.
If the input file contains interforce and/or adhesiveforce in the section field, the commands
interforce and adhesiveforce in the script redefine the parameters $G$, $\psi_{0}$ and $\rho_{0}$.

With an additional argument of type solute, the file defining the solute state is read in
(see write\_restart command). For solute transport, the information of the input data file must
be organized into seven sections: relaxation\_time, inlet, outlet, reaction, field, concentration and solute\_modes.

The relaxation\_time section is defined as follows:\\
relaxation\_time\\
$\tau_{\mathrm{s}}$

The inlet and/or outlet sections are defined as follows:\\
inlet\\
wall $x_{0}$ $x_{1}$ label style arguments\\
outlet\\
wall $x_{0}$ $x_{1}$ label style arguments\\
The details are explained in the description of the commands inlet/outlet and inlet\_solute/outlet\_solute.

The reaction section is defined as follows:\\
reaction\\
surface $k_{\mathrm{r}}$ $C_{\mathrm{s}}$ style $b_{\mathrm{max}}$\\
The various parameters are introduced in the description of the command surface\_reaction. This
section must precede the section concentration.

The field section is defined as follows:\\
field\\
intersolute $G$ $\psi_{0}$ $C_{0}$\\
The parameters are introduced in the description of the command intersolute.

The concentration section is defined as follows:\\
concentration\\
$x$ $y$ $C$ $b$\\
$x$ and $y$ are the coordinates of the lattice site and $C$ its solute concentration.
In the case of surface reaction, $b$ is the cumulated mass on nodes belonging to the solid surface;
for the fluid phase $b$ vanishes. This quantity is omitted in the absence of surface
reaction. For clarity, we suggest to set equal to zero the concentration of the solid 
nodes not lying on the surface. This section must follow the section reaction.

The solute\_modes section is defined as follows:\\
solute\_modes\\
$x$ $y$ $g_{1}$ $g_{2}$ $g_{3}$ $g_{4}$\\
$x$ and $y$ are the coordinates of the lattice site. On a d2q4 lattice, 
$g_{1}$,$\dots$,$g_{4}$ are the solute distribution functions. For clarity, we suggest
to set all the distribution functions equal to zero for the solid nodes not lying on 
the surface.

Also for solute transport, the section field must follow the section relaxation\_time.
In a script, if the solute command is used after the read\_data command, the relaxation time is modified.
If the input file contains intersolute in the section field, the command intersolute in
the script redefines the parameters $G$, $\psi_{0}$ and $C_{0}$. The same holds for the
section reaction and the corresponding command surface\_reaction.

In general, the name of a section must be separated by at least one new line character
from the data. Different numeric values must be separated by at least one white space
character or tabular character. For example, the following line is wrong:\\
relaxation\_time $1$\\
The following lines are instead read correctly:\\
\verb1      1density

$1$\hspace{1cm}$0$\hspace{1.5cm} $84.111$

The restart file of a previous simulation can be used as input file
of a subsequent simulation.

\underline{Related commands:}

inlet/outlet inlet\_momentum/outlet\_momentum surface\_reaction aveforce aveacceleration
interforce adhesiveforce intersolute write\_restart
\vspace{0.5cm}

\textbf{\underline{fluids command}}

\underline{Syntax:}

fluids(tau1,tau2,alpha1,alpha2)
\begin{itemize}
\item tau1=relaxation time fluid1
\item tau2=relaxation time fluid2
\item alpha1=parameter controlling density ratio
\item alpha2=parameter controlling density ratio
\end{itemize}

\underline{Examples:}

\&fluids(1,2);\\
\&fluids(0.626,2.5,4/9,0.8);\\
\&fluids(1,1,0,0.998);

\underline{Description:}

With this command it is possible to simulate systems containing two substances. Details on the formalism can be found in 
Refs.~\cite{book1,martys,theta}. If the parameters $\alpha_{\sigma}$ are omitted, 
they are set to $\alpha_{\sigma}=4/9$ for both fluid components $\sigma=1,2$. In this case, the equilibrium distribution functions
are given by Eq.~\ref{eq:eq}, but using a different velocity. In general, the equilibrium distribution functions take
the form \cite{chibbaro,succi1}
\begin{equation*}
f^{\sigma,\mathrm{eq}}_{i}(\rho_{\sigma},\bm{u}_{\sigma}^{\mathrm{eq}})=w_{i}\rho_{\sigma}(\bm{r},t)\Big[A^{\sigma}_{i}
+3\bm{e}_{i}\cdot\bm{u}_{\sigma}^{\mathrm{eq}}+\frac{9}{2}(\bm{e}_{i}\cdot\bm{u}_{\sigma}^{\mathrm{eq}})^{2}
-\frac{3}{2}\bm{u}_{\sigma}^{\mathrm{eq}}\cdot\bm{u}_{\sigma}^{\mathrm{eq}}\Big]\quad i=0,\dots,8\ .
\end{equation*}
The factors $w_{i}$ are the usual weights for the d2q9 lattice (see Eq.~\ref{eq:eq}). $\sigma$ designates the fluid component. $\rho_{\sigma}$ 
is the local density; $\bm{u}_{\sigma}^{\mathrm{eq}}$ is the equilibrium velocity determined as explained in Refs.~\cite{book1,martys,theta}. 
The coefficients $A^{\sigma}_{i}$ are given by $A^{\sigma}_{0}=(9/4)\alpha_{\sigma}$ for $i=0$; $A^{\sigma}_{i}=(9/5)(1-\alpha_{\sigma})$ for $i=1-8$. 
The sound speed of the $\sigma$-th component becomes $(c^{\sigma}_{\mathrm{s}})^{2}=(3/5)(1-\alpha_{\sigma})$.
\vspace{0.5cm}

\textbf{\underline{solute command}}

\underline{Syntax:}

solute(taus)
\begin{itemize}
\item taus=relaxation time for solute transport
\end{itemize}

\underline{Example:}

\&solute(1);

\underline{Description:}

This command enables solute transport for one single species. The relaxation time for
solute $\tau_{\mathrm{s}}$ determines the diffusion coefficient
\begin{equation*}
D=\frac{1}{2}\Big(\tau_{\mathrm{s}}-\frac{1}{2}\Big)\ .
\end{equation*}
Solute transport is accounted for using the distribution functions $g_{i}$ defined on a
d2q4 lattice \cite{d2q4}. The dynamics develops through two steps, namely streaming and 
collision (see Fig.~\ref{fig:algorithm}). As for fluid flow, the evolution is governed
by the BGK equation \cite{bgk}
\begin{equation*}
g_{i}(\bm{r}+\bm{e}_{i}\Delta t,t+\Delta t)=g_{i}(\bm{r},t)
-\frac{1}{\tau_{\mathrm{s}}}\big[g_{i}(\bm{r},t)-g_{i}^{\mathrm{eq}}(C,\bm{u})\big]\ ,\quad
i=1,\dots,4\ .
\end{equation*}
In the collision term, the equilibrium functions are of the form \cite{d2q4}
\begin{equation}
g^{\mathrm{eq}}_{i}(C,\bm{u})=\frac{1}{4}C[1+2\bm{e}_{i}\cdot\bm{u}]\ ,\quad i=1,\dots,4\ .
\label{eq:ueq}
\end{equation}
The concentration $C$ is given by 
\begin{equation*}
C(\bm{r},t)=\sum_{i=1}^{4}g_{i}(\bm{r},t)\ .
\end{equation*}
$\bm{u}$ is the velocity of the solvent. In the multicomponent case, it is given by \cite{u_fluids}
\begin{equation}
\bm{u}(\bm{r},t)=\Big(\sum_{\sigma}\Big[\rho_{\sigma}(\bm{r},t)\bm{u}_{\sigma}(\bm{r},t)
+\frac{1}{2}\bm{F}_{\sigma}(\bm{r},t)\Big]\Big)\bigg/\sum_{\sigma}\rho_{\sigma}(\bm{r},t)\ .
\label{eq:u_solvent}
\end{equation}
$\rho_{\sigma}$ designates the local density of the fluid components; $\bm{F}_{\sigma}$
is the force experienced by the fluid component denoted by $\sigma$. The fluid
velocity of the $\sigma$-th component is 
\begin{equation*}
\bm{u}_{\sigma}(\bm{r},t)=\frac{1}{\rho_{\sigma}(\bm{r},t)}
\sum_{i}f_{i}^{\sigma}(\bm{r},t)\bm{e}_{i}\ .
\end{equation*}
More details on the theory for solute transport can be found in 
Refs.~\cite{d2q4,reaction_pre,dissolution,crystal,snow}.

At the surface of the solid phase, the bounce-back rule is applied, unless the surface reaction is
enabled.

\underline{Related commands:}

inlet\_solute/outlet\_solute surface\_reaction boundary\_style concentration0 
\vspace{0.5cm}

\textbf{\underline{inlet/outlet commands}}

\underline{Syntax:}

inlet(wall,$x_{0}$,$x_{1}$,label1,label2)\\
outlet(wall,$x_{0}$,$x_{1}$,label1,label2)
\begin{itemize}
\item wall=xhi,xlo,yhi,ylo is the wall of the simulation domain where the opening lies
\item $x_{0}<x_{1}=$coordinates of the extremities of the opening ($y_{0}<y_{1}$ for xhi and xlo)
\item label1=number or string referencing the opening
\item label2=number or string referencing the opening for the second fluid
\end{itemize}

\underline{Examples:}

\&inlet(``xhi'',5,10,11);\\
\&outlet(``ylo'',3,50,''out1'');\\
\&inlet(``xlo'',1,49,''in1'',''in2'');\\
\&inlet(``ylo'',1,49,11,12);

\underline{Description:}

\begin{figure}[t]
\includegraphics[height=3.5cm]{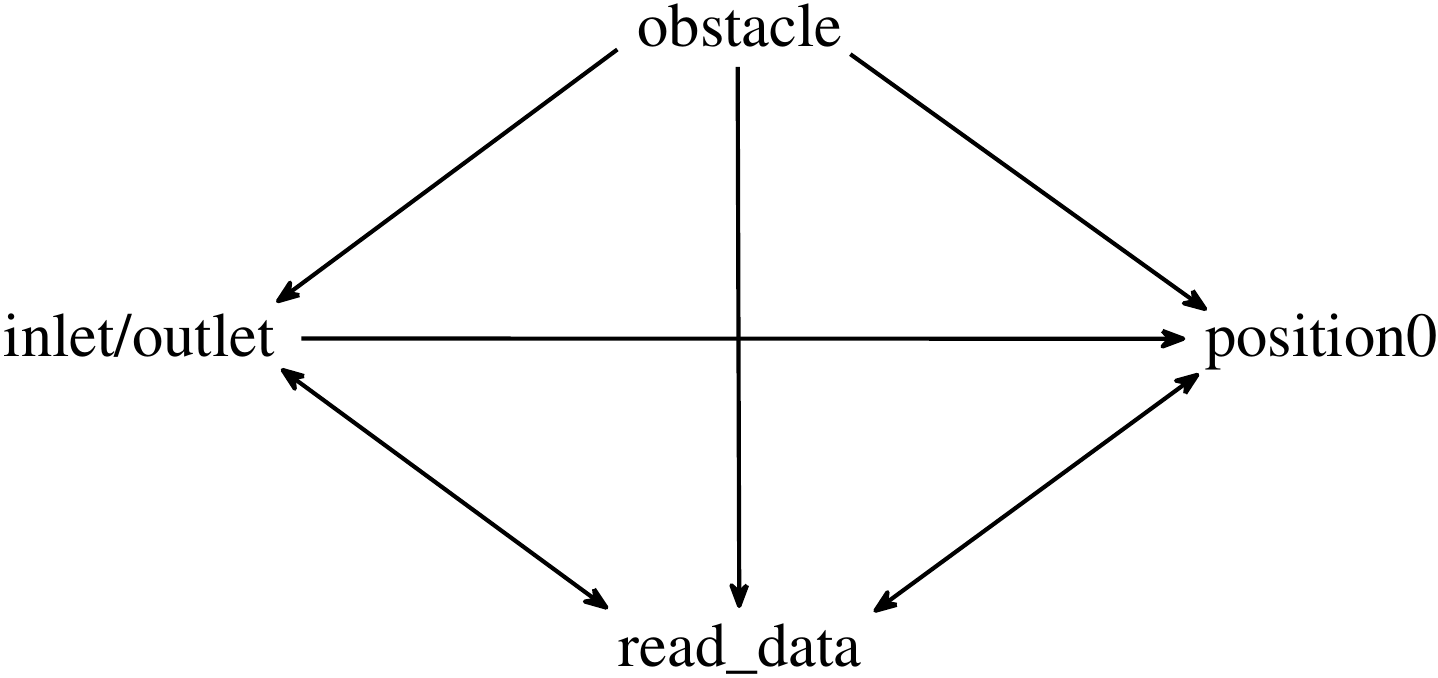}
\caption{
Hierarchy among the commands of initialization. A command points to another one when
it overrides its previous definitions.
\label{fig:hierarchy}}
\end{figure}

With these commands the location of the inlet and/or outlet openings is defined.
They override the previous settings via the read\_data command as illustrated in
Fig.~\ref{fig:hierarchy}.

In the multicomponent case, it is necessary to give two labels, one for every fluid component.

\underline{Related commands:}

inlet\_momentum/outlet\_momentum position0 momentum0
\vspace{0.5cm}

\textbf{\underline{inlet\_momentum/outlet\_momentum commands}}

\underline{Syntax:}

inlet\_momentum(label,style,arguments)\\
outlet\_momentum(label,style,arguments)
\begin{itemize}
\item label specifies the opening previously defined with the commands inlet/outlet
\item style=poiseuille, uniform, dirichlet and gradient are available for both inlet and outlet
\item arguments depend on the style of the opening:
\begin{itemize}
\item If the opening is of style poiseuille, it is necessary to specify the maximal velocity $v_{\mathrm{max}}$
and the initial density $\rho$ at the opening nodes.
\item For the uniform style, it is
necessary to give the velocity and the initial density $\rho$ of the opening nodes.
\item dirichlet means constant pressure. In this case, it is necessary to specify the initial density
$\rho$ to hold constant at the opening nodes.
\item For the gradient style, it is necessary to enter the initial density of the opening nodes.
\end{itemize}
\end{itemize}

\underline{Examples:}

\&inlet\_momentum(``in1'',''poiseuille'',0.07,24);\\
\&inlet\_momentum(100,''uniform'',0.05,24);\\
\&inlet\_momentum(``in3'',''dirichlet'',24.25);\\
\&outlet\_momentum(``out3'',''dirichlet'',24);\\
\&outlet\_momentum(``out2'',''gradient'',24);

\underline{Description:}

These commands determine how to update the momentum at the nodes of the inlet
and/or outlet openings. More precisely, for an inlet opening on the wall xlo,
after streaming and the treatment of collisions with solid boundaries the velocity modes $f_{1}$,
$f_{5}$ and $f_{8}$ remain undetermined. For the styles poiseuille and uniform, they
are updated according to the following rules \cite{inlet}:
\begin{eqnarray*}
&&\rho=\frac{1}{1-v_{x}}[f_{0}+f_{2}+f_{4}+2(f_{3}+f_{6}+f_{7})]\ ,\\
&&f_{1}=f_{3}+\frac{2}{3}\rho v_{x}\ ,\\
&&f_{5}=f_{7}-\frac{1}{2}(f_{2}-f_{4})+\frac{1}{6}\rho v_{x}\ ,\\
&&f_{8}=f_{6}+\frac{1}{2}(f_{2}-f_{4})+\frac{1}{6}\rho v_{x}\ .
\end{eqnarray*}
$v_{x}$ is the velocity imposed at the inlet node. For the poiseuille style we have
\begin{equation*}
v_{x}(y)=\frac{4v_{\mathrm{max}}}{(y_{1}-y_{0}+1)^{2}}\Big[\Big(\frac{y_{1}-y_{0}+1}{2}\Big)^{2}
-\Big(\frac{2y-y_{1}-y_{0}}{2}\Big)^{2}\Big]\quad\quad\text{for }y=y_{0},\dots,y_{1}\ .
\end{equation*}
The velocity $v_{x}$ is of course constant in the uniform case.
The collision step is then performed including also the nodes of the opening \cite{inlet}.
For the other walls, the formulas are given in Appendix \ref{ap:inout}.

For the style dirichlet, the pressure $P$ is kept constant at the inlet and/or outlet openings
($P=c_{\mathrm{s}}^{2}\rho$). Let us consider the wall xhi for example. After streaming, the
velocity $v_{x}$ and the populations $f_{3}$, $f_{6}$, $f_{7}$ are unknown. They
are determined as follows:
\begin{eqnarray*}
&&v_{x}=\frac{f_{0}+f_{2}+f_{4}+2(f_{1}+f_{5}+f_{8})}{\rho}-1\ ,\\
&&f_{3}=f_{1}-\frac{2}{3}\rho v_{x}\ ,\\
&&f_{6}=f_{8}+\frac{1}{2}(f_{4}-f_{2})-\frac{1}{6}\rho v_{x}\ ,\\
&&f_{7}=f_{5}-\frac{1}{2}(f_{4}-f_{2})-\frac{1}{6}\rho v_{x}\ .
\end{eqnarray*}
Then, the collision step is performed also for the nodes of this opening \cite{inlet}.
The openings on the other walls are treated as explained in Appendix \ref{ap:inout}.

For the style gradient, after the collision step, the velocity modes are updated
with the values of their neighbors. For example, for an outlet opening on the xhi wall,
we have $f_{i}(N_{x},y)=2f_{i}(N_{x}-1,y)-f_{i}(N_{x}-2,y)$ for $i=0,\dots,8$.

With two substances (see fluids command), it is necessary to specify how to update the momentum
for every fluid component: the first label in the inlet/outlet command refers to fluid1 and the second
one to fluid2.

Warning: The thermo information is not correct with the gradient style (see thermo command).

\underline{Related commands:}

inlet/outlet thermo
\vspace{0.5cm}

\textbf{\underline{inlet\_solute/outlet\_solute commands}}

\underline{Syntax:}

inlet\_solute(label,style,arguments)\\
outlet\_solute(label,style,arguments)
\begin{itemize}
\item label specifies the opening previously defined with the commands inlet/outlet
\item style=flux, dirichlet and adiabatic are available for both inlet and outlet
\item arguments depend on the style of the opening:
\begin{itemize}
\item If the opening is of style flux, it is necessary to specify the velocity $v$ and the solute 
concentration $C_{0}$.
\item For the dirichlet style, the solute concentration is kept constant. The initial and imposed solute 
concentration is an input parameter.
\item For the adiabatic style, the initial solute concentration must be specified.
\end{itemize}
\end{itemize}

\underline{Examples:}

\&inlet\_solute(``in1'',''flux'',0.01,1.1);\\
\&inlet\_solute(99,''dirichlet'',1);\\
\&outlet\_soltute(``out1'',''dirichlet'',1.1);\\
\&outlet\_solute(``out2'',''adiabatic'',1);

\underline{Description:}

These commands determine how to update the solute concentration at the nodes of the inlet
and/or outlet openings. As for fluid flow, after the streaming step some distribution functions
remain undetermined. For example, for an inlet opening on the wall xlo, the distribution function
$g_{1}$ is not defined (transport occurs on a d2q4 lattice, see solute command). For the style flux, 
it is determined according to the rules \cite{book1}:
\begin{equation*}
C=\frac{vC_{0}+DC_{1}}{D+v}\quad\text{and}\quad g_{i}=g_{i}^{\mathrm{eq}}\ ,\quad i=1,\dots,4\ .
\end{equation*}
$C$ is the solute concentration at the inlet node. $C_{0}$ is the concentration of the solution
injected into the system and $v$ is the associated velocity. $C_{1}$ is the solute concentration of the nearest 
node inside the simulation domain. The collision step is applied also to the nodes of this opening.
The formulas for the other walls are given in Appendix \ref{ap:inout_solute}.

For the style dirichlet, the concentration $C$ is kept constant at the inlet and/or outlet openings. 
For example, for the wall xhi, after streaming, the distribution $g_{3}$ is unknown 
(transport on a d2q4 lattice, see solute command). It is determined according to the following 
rules \cite{book1}:
\begin{equation*}
C=\bar{C}\quad\text{and}\quad
g_{3}=\bar{C}-C'\ .
\end{equation*}
$C$ is the solute concentration at the opening. $\bar{C}$ is the desired solute concentration; $C'$ is the 
actual value immediately after the streaming step. The collision step is then performed also for the nodes of this opening.
The formulas for the other walls can be found in Appendix \ref{ap:inout_solute}.

For the style adiabatic, we impose a vanishing flux at the nodes of the openings. For example, for an outlet opening 
on the xhi wall, we simply have $g_{3}=g_{1}$. The collision step is then applied also for these lattice sites.

With two substances (see inlet/outlet commands), it is possible to use the label of any fluid component.

\underline{Related commands:}

solute inlet/outlet
\vspace{0.5cm}

\textbf{\underline{boundary\_style command}}

\underline{Syntax:}

boundary\_style(style,argument)
\begin{itemize}
\item style=noslip,slip,mixed,hybrid
\item argument=reflection coefficient for the mixed style (argument$\leq1$)
\end{itemize}

\underline{Examples:}

\&boundary\_style(``slip'');\\
\&boundary\_style(``noslip'');\\
\&boundary\_style(``mixed'',0.5);\\
\&boundary\_style(``hybrid'');

\underline{Description:}

With this command it is specified how to treat the collisions at the
solid boundaries. This command must precede the obstacle command, otherwise the simulation
is interrupted with an error message.

In the absence of this command all boundaries are for sure periodic: in the restart
file the style of the boundaries is indicated as periodic, even with inlet/outlet openings.

The noslip style is suited for describing rough surfaces. The velocity
modes reaching an obstacle are simply bounced back \cite{book1,book2}.
\begin{figure}[t]
\begin{center}
\includegraphics[width=6.5cm]{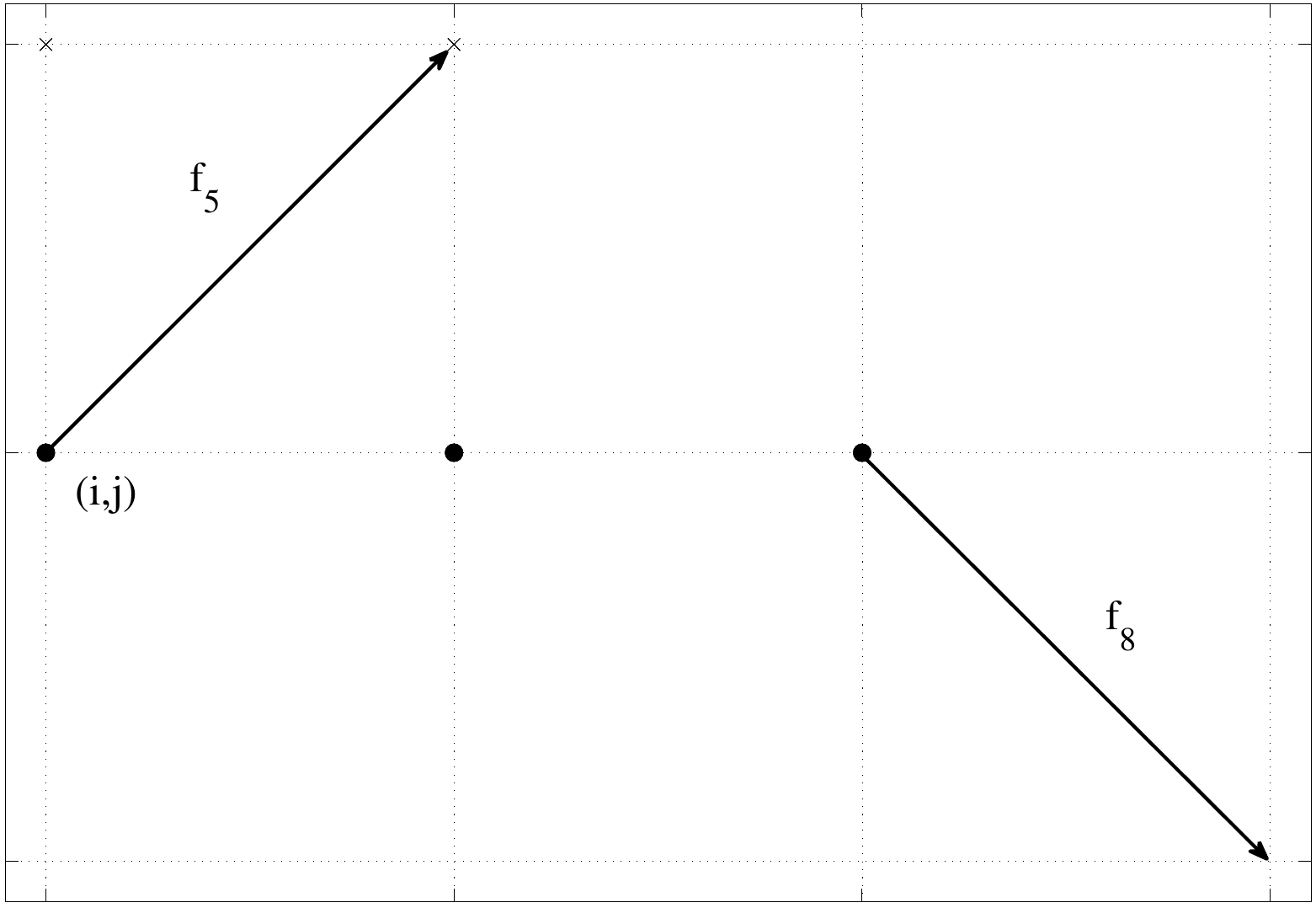}\hspace{3.5cm}
\includegraphics[height=6cm]{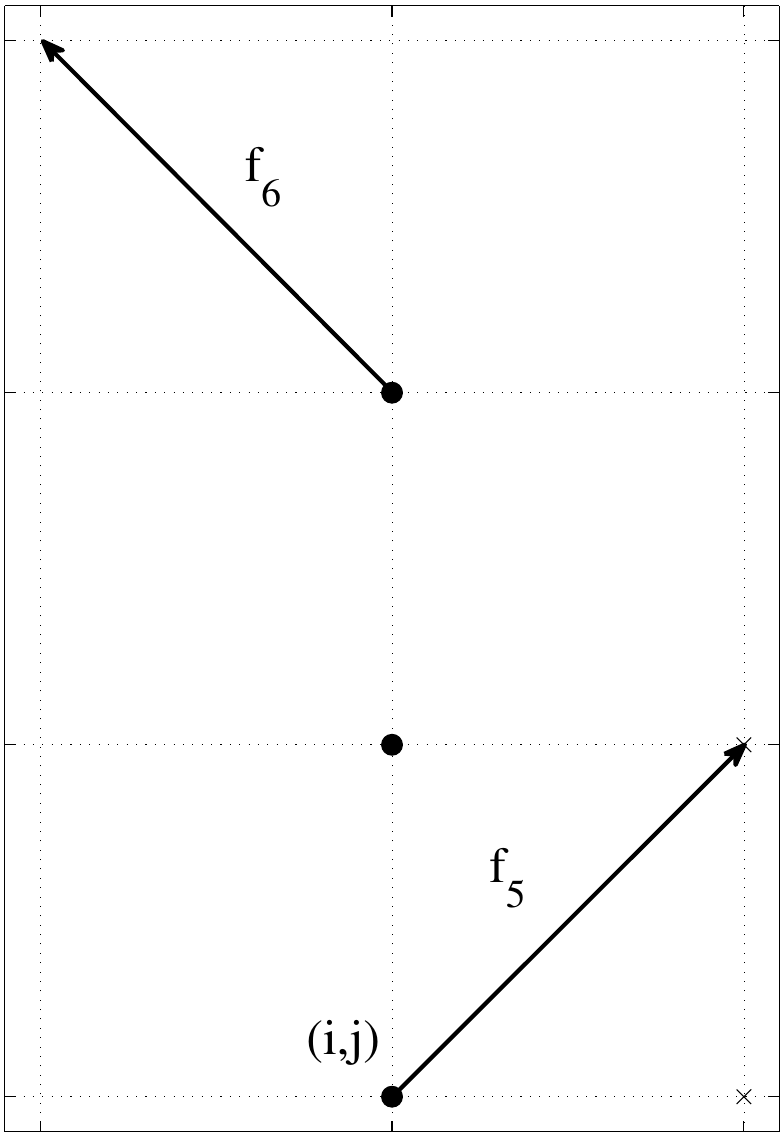}
\end{center}
\caption{Schematic representation of the collision rule for the velocity mode
$f_{5}$ with the boundary style slip (cf.~main text). Filled circles represent
the fluid phase while crosses obstacles.}
\label{fig:slip}
\end{figure}

The slip style is useful for simulating smooth surfaces (mirror reflection).
The velocity modes $f_{1}$, $f_{2}$, $f_{3}$ and $f_{4}$ are still bounced back
when they meet an obstacle. The diagonal modes are treated differently. Let us
consider for example the velocity mode $f_{5}$ at node $(i,j)$. If the node
$(i+1,j+1)$ is not an obstacle, the velocity mode $f_{5}$ propagates freely.
Otherwise, for the arrangements illustrated in Fig.~\ref{fig:slip}, we
apply the rules
\begin{equation*}
f_{8}(i+2,j)=f_{5}(i,j)\quad\text{or}\quad
f_{6}(i,j+2)=f_{5}(i,j)\ .
\end{equation*}
In all the remaining cases, the velocity mode is bounced back. Similar
rules apply for the other diagonal modes. Care is taken in order to avoid that
velocity modes cross inlet/outlet openings.

The mixed style is a combination of the slip and noslip rules. When a velocity mode
hits a solid site, a fraction $R$ is propagated forward (slip rule) and the remaining
fraction $1-R$ backward (noslip rule).

The mode hybrid allows to use more than one style for the boundaries. The type
of a solid node is specified via the command obstacle with additional arguments (the extra arguments
double jointly with the fluids command). In simulations, the collision is treated
according to the style of the target solid node.

\underline{Related command:}

obstacle
\vspace{0.5cm}

\textbf{\underline{obstacle command}}

\underline{Syntax:}

obstacle(rectangle,$x_{0}$,$x_{1}$,$y_{0}$,$y_{1}$,style1,argument,style2,argument)\\
obstacle(circle,$x_{\mathrm{c}}$,$y_{\mathrm{c}}$,$r$,style1,argument,style2,argument)
\begin{itemize}
\item shape=rectangle,circle
\item $x_{0}<x_{1}$=lower and higher $x$ coordinates for the
  rectangle. The same holds for $y_{0}$ and $y_{1}$
\item $x_{\mathrm{c}}$,$y_{\mathrm{c}}$=coordinates of the center of the circle
\item $r$=radius of the circle
\item style1,style2=noslip,slip,mixed
\item argument=reflection coefficient for the mixed style (argument$\leq1$)
\end{itemize}

\underline{Examples:}

\&obstacle(``rectangle'',4,10,5,5);\\
\&obstacle(``circle'',5,5,10);\\
\&obstacle(``rectangle'',0,0,0,50,''slip'');\\
\&obstacle(``circle'',20,10,5,''noslip'');\\
\&obstacle(``rectangle'',0,60,0,0,''noslip'',''noslip'');\\
\&obstacle(``rectangle'',2,7,12,17,''mixed'',0.3);\\
\&obstacle(``rectangle'',10,20,10,30,''mixed'',0.7,''mixed'',0.5);

\underline{Description:}

This command defines the lattice sites that will be considered as occupied
by a solid phase. This command overrides any previous definition of these
lattice sites via the read\_data, inlet/outlet and/or postion0 commands (see Fig.~\ref{fig:hierarchy}).
This command must follow the boundary\_style command, otherwise the simulation
is interrupted with an error message.

If the boundary style is hybrid, the command takes extra arguments
specifying one of the available methods for the treatment of collisions.
In the multicomponent case, the additional arguments double.

With the mixed style, at most $99$ different reflection coefficients can be
defined for the first substance in the multicomponent case.

\underline{Related commands:}

read\_data boundary\_style position0
\vspace{0.5cm}

\textbf{\underline{surface\_reaction command}}

\underline{Syntax:}

surface\_reaction(kr,Cs,style,b0,bmax)
\begin{itemize}
\item kr=reaction-rate constant
\item Cs=saturated concentration
\item style=growth
\item b0,bmax=initial and threshold value for cumulated mass at solid-liquid interface
\end{itemize}

\underline{Example:}

\&surface\_reaction(0.25,1,''growth'',0.5,1);

\underline{Description:}

This command introduces a surface reaction for solute transport. At the reactive boundary,
the solute concentration $C$ satisfies the macroscopic equation 
\cite{d2q4,reaction_pre,dissolution,crystal,snow}
\begin{equation*}
D\frac{\partial C}{\partial n}=k_{\mathrm{r}}(C-C_{\mathrm{s}})\ ,
\end{equation*}
where $D$ is the diffusion coefficient, $C_{\mathrm{s}}$ is the saturated concentration and
$k_{\mathrm{r}}$ is the reaction-rate constant. After the streaming step for solute transport,
at the solid surface the incoming modes into the fluid phase are undetermined. The unknown
quantities are updated according to the following rules \cite{d2q4}:
\begin{equation*}
C=\frac{2g_{\chi(i)}+k_{\mathrm{r}}C_{\mathrm{s}}}{k_{\mathrm{r}}+1/2}\quad\quad\text{and}\quad\quad
g_{i}=-g_{\chi(i)}+\frac{1}{2}C\ .
\end{equation*}

The distribution functions $g_{i}$ describe solute transport (see solute command). $i'=\chi(i)$
is the index for the opposite direction associated with that of $i$. More details on the
implementation can be found in Appendix \ref{ap:d2q4}. The subsequent  collision step is
performed also for the nodes at the solid-liquid interface. After the collision step, at the
solid boundaries mass precipitates. The mass deposited on a solid node is updated as follows 
\cite{crystal,snow}:
\begin{equation*}
b(t+\Delta t)=b(t)+\Delta b\quad\text{with}\quad \Delta b=k_{\mathrm{r}}(C-C_{\mathrm{s}})\ ,
\end{equation*}
if $\Delta b>0$. The initial mass on solid nodes is $b_{0}$. The amount $\Delta b$ is
subtracted from the incoming modes in the fluid phase \cite{snow}. Solute transport
in the solid phase is neglected. For surface growth, when for a solid node
$b(t+\Delta t)$ exceeds $b_{\mathrm{max}}$, one of its direct neighbors in the fluid phase
becomes a solid node; it is chosen randomly and every occurrence is equiprobable.

\underline{Related command:}

solute
\vspace{0.5cm}

\textbf{\underline{position0 command}}

\underline{Syntax:}

position0(rectangle,$x_{0}$,$x_{1}$,$y_{0}$,$y_{1}$)\\
position0(circle,$x_{\mathrm{c}}$,$y_{\mathrm{c}}$,$r$)
\begin{itemize}
\item shape=rectangle,circle
\item $x_{0}<x_{1}$=lower and higher $x$ coordinates for the rectangle. The
  same holds for $y_{0}$ and $y_{1}$
\item $x_{\mathrm{c}}$,$y_{\mathrm{c}}$=coordinates of the center of the circle
\item $r$=radius of the circle
\end{itemize}

\underline{Examples:}

\&position0(``rectangle'',8,9,11,20);\\
\&position0(``circle'',5,7,11);

\underline{Description:}

This command defines regions of rectangular and/or circular shape for
which the user specifies the starting populations via the momentum0
command.

The previous definitions of the lattice sites via the read\_data command are
modified. Inlet/outlet nodes and/or obstacle nodes are not reinitialized (see Fig.~\ref{fig:hierarchy}).

This command must precede the command momentum0, otherwise the simulation
is interrupted with an error message.

\underline{Related command:}

momentum0

\vspace{0.5cm}

\textbf{\underline{momentum0 command}}

\underline{Syntax:}

momentum0(mode,value,fluid)
\begin{itemize}
\item mode=e0,e1,e2,$\dots$ velocity mode
\item value=population value
\item fluid=1,2 indicates the substance
\end{itemize}

\underline{Examples:}

\&momentum0(``e0'',4);\\
\&momentum0(``e7'',1);\\
\&momentum0(``e1'',4,1);

\underline{Description:}

This command allows to set the populations of the lattice sites selected
via the position0 command. If two commands are used with the same mode,
the populations are summed. Every momentum0 command refers to the last region
defined with the position0 command: regions with different densities can be
so defined. Negative population values are subtracted.

This command must be used after the command position0, otherwise the simulation
is interrupted with an error message. The other priority relations are illustrated
in Fig.~\ref{fig:hierarchy}.

In the multicomponent case, it is necessary to specify the fluid component with an
extra argument: 1 for fluid1, 2 for fluid2.

\underline{Related commands:}

inlet/outlet position0
\vspace{0.5cm}

\textbf{\underline{concentration0 command}}

\underline{Syntax:}

concentration0(value)
\begin{itemize}
\item value=concentration value
\end{itemize}

\underline{Example:}

\&concentration0(1.3);

\underline{Description:}

This command allows to define the solute concentration for the lattice sites selected via
the position0 command. Every concentration0 command refers to the last region defined with
the position0 command: regions with different concentrations can be so defined. Negative
concentration values are subtracted.

This command must be used after the command position0, otherwise the simulation is 
interrupted with an error message.

\underline{Related command:}

solute
\vspace{0.5cm}

\textbf{\underline{aveforce command}}

\underline{Syntax:}

aveforce(mode,value,fluid)
\begin{itemize}
\item mode=e0,e1,e2,$\dots$ velocity mode
\item value=intensity of the force
\item fluid=1,2 indicates the substance
\end{itemize}

\underline{Examples:}

\&aveforce("e1",$3.11$*$10$**$-3$);\\
\&aveforce("e1",$4.33$*$10$**$-3$,1);

\underline{Description:}

With this command an average force is applied to all lattice sites
occupied by the fluid phase. If two commands are used with the same mode,
the components of the force vector are summed. This command must be used
after the lattice or fluids commands.

In the multicomponent case, it is necessary to specify the fluid component with an
extra argument: 1 for fluid1, 2 for fluid2.

\underline{Related command:}

aveacceleration

\vspace{0.5cm}

\textbf{\underline{aveacceleration command}}

\underline{Syntax:}

aveacceleration(mode,value,fluid)
\begin{itemize}
\item mode=e0,e1,e2,$\dots$ velocity mode
\item value=intensity of the acceleration
\item fluid=1,2 indicates the substance
\end{itemize}

\underline{Examples:}

\&aveacceleration("e1",$2.22$*$10$**$-4$);\\
\&aveacceleration("e3",$1.55$*$10$**$-4$,2);

\underline{Description:}

With this command an average acceleration is added to all lattice sites
occupied by the fluid phase. If two commands are used with the same mode,
the components of the vector acceleration are summed. This command must be used
after the lattice or fluids commands.

In the multicomponent case, it is necessary to specify the fluid component with an
extra argument: 1 for fluid1, 2 for fluid2.

\underline{Related command:}

aveforce

\vspace{0.5cm}

\textbf{\underline{interforce command}}

\underline{Syntax:}

interforce($G$,$\psi_0$,$\rho_0$)
\begin{itemize}
\item $G$=interaction strength
\item $\psi_0$=factor
\item $\rho_0$=average density
\end{itemize}

\underline{Examples:}

\&interforce(-120,4,200);\\
\&interforce(0.9);

\underline{Description:}

This command introduces van der Waals forces between nearest-neighbor
lattice sites occupied by the fluid phase. The force is computed
with the formula
\begin{equation*}
\bm{F}(x,y)=-G\psi(x,y)\sum_{i=1}^{8}w_{i}
\psi(\bm{r}+\bm{e}_{i}\Delta t)\bm{e}_{i}\ .
\end{equation*}
We use the notation $\bm{r}=(x,y)$ for a two dimensional lattice; time
dependence omitted for clarity. The constants $w_{i}$ are defined by the 
lattice style. The function $\psi(x,y)$ is defined as
\begin{equation*}
\psi(x,y)=\psi_{0}e^{-\rho_{0}/\rho(x,y)}\ .
\end{equation*}

In the multicomponent case, the above formula becomes \cite{book1,martys,theta}
\begin{equation*}
\bm{F}_{\sigma}(x,y)=-G\rho_{\sigma}(x,y)\sum_{i=1}^{8}w_{i}
\rho_{\bar{\sigma}}(\bm{r}+\bm{e}_{i}\Delta t)\bm{e}_{i}\quad\quad\sigma=1,2\ .
\end{equation*}
Given $\sigma$, the other fluid component is denoted as $\bar{\sigma}$. Of course,
the arguments $\psi_{0}$ and $\rho_{0}$ must be omitted in this case.

Warning: The thermo information is not correct when this command is used (see thermo command).

\underline{Related command:}

thermo

\vspace{0.5cm}

\textbf{\underline{adhesiveforce command}}

\underline{Syntax:}

adhesiveforce($G$,$\psi_0$,$\rho_0$,fluid)
\begin{itemize}
\item $G$=interaction strength
\item $\psi_0$=factor
\item $\rho_0$=average density
\item fluid=1,2 indicates the substance
\end{itemize}

\underline{Examples:}

\&adhesiveforce(-200,4,200);\\
\&adhesiveforce(-0.2,1);

\underline{Description:}

This command introduces adhesive forces between lattice sites occupied
by the fluid phase in the proximity of solid boundaries. The force
is computed with the formula
\begin{equation*}
\bm{F}(x,y)=-G\psi(x,y)\sum_{i=1}^{8}w_{i}
s(\bm{r}+\bm{e}_{i}\Delta t)\bm{e}_{i}\ .
\end{equation*}
We use the notation $\bm{r}=(x,y)$ for a two dimensional lattice; time
dependence omitted for clarity. The constants $w_{i}$ are defined by the 
lattice style. The function $\psi(x,y)$ is given by
\begin{equation*}
\psi(x,y)=\psi_{0}e^{-\rho_{0}/\rho(x,y)}\ .
\end{equation*}
The function $s(x,y)$ is defined as
\[
s(x,y)=
\left\{\begin{array}{l}
1\quad\text{if site $(x,y)$ solid}\\
0\quad\text{if site $(x,y)$ fluid}
\end{array}\ .
\right.\]

In the multicomponent case, it is necessary to specify the fluid component with an
extra argument: 1 for fluid1, 2 for fluid2. The force is evaluated with the formula
\cite{book1,martys,theta}
\begin{equation*}
\bm{F}_{\sigma}(x,y)=-G_{\sigma}\rho_{\sigma}(x,y)\sum_{i=1}^{8}w_{i}
s(\bm{r}+\bm{e}_{i}\Delta t)\bm{e}_{i}\quad\quad\sigma=1,2\ .
\end{equation*}
Of course, the arguments $\psi_{0}$ and $\rho_{0}$ must be omitted in this case.
\vspace{0.5cm}

\textbf{\underline{intersolute command}}

\underline{Syntax:}

intersolute($G$,$\psi_0$,$C_0$)
\begin{itemize}
\item $G$=interaction strength
\item $\psi_0$=factor
\item $C_0$=average concentration
\end{itemize}

\underline{Example:}

\&intersolute($-5$*$10$**$-3$,1,$5$*$10$**$-3$);

\underline{Description:}

With this command it is possible to introduce interfaces for solute transport.
In analogy with multiphase fluid flow, the mechanism consists in adding to the
velocity $\bm{u}$ in Eq.~\ref{eq:ueq} a term of the form 
$\tau_{\mathrm{s}}\bm{F}(\bm{r},t)/C(\bm{r},t)$. The function
$\bm{F}$ is obtained by
\begin{equation*}
\bm{F}(x,y)=-G\psi(x,y)\sum_{i=1}^{4}\psi(\bm{r}
+\bm{e}_{i}\Delta t)\bm{e}_{i}\ .
\end{equation*}
We use the notation $\bm{r}=(x,y)$ for a two dimensional lattice; time dependence
omitted for clarity. The function $\psi(x,y)$ is defined as
\begin{equation*}
\psi(x,y)=\psi_{0}e^{-C_{0}/C(x,y)}\ .
\end{equation*}

\underline{Related commands:}

solute interforce

\vspace{0.5cm}

\textbf{\underline{thermo command}}

\underline{Syntax:}

thermo($N_{\mathrm{f}}$)
\begin{itemize}
\item $N_{\mathrm{f}}$=print general information on screen every this many timesteps
\end{itemize}

\underline{Example:}

\&thermo(50);

\underline{Description:}

When this command is used, general information is printed on screen during
the simulation. Mass is the weight of the overall fluid phase of the system. It is
obtained by summing the density of the lattice sites occupied by the fluid phase.
Momentum returns the $x$ and $y$ components of the overall momentum vector.
Mass\_eq is the equilibrium mass of the overall fluid phase; it is calculated
from the equilibrium densities. At $t=0$, this information refers to the state of the system
before the routine of streaming (see Fig.~\ref{fig:algorithm}). For $t>0$, it is given
the state of the system after the collision step (see Fig.~\ref{fig:algorithm}).
At the beginning and at the end of a run the local time is printed.
At start, it is also indicated if the program runs in serial or parallel mode. For example,
$1\times 1$ is used for serial; $2\times 1$ means that the $x$ side of the simulation domain is
partitioned between $2$ processors.

The basic computations for the thermo information are performed in the collision step,
thus without additional loops. When the command interforce is used, the contribution of
this force to the velocity is not computed correctly because all the densities are updated
only at the end of the loops. Similarly, with the gradient style for inlet/outlet openings,
the thermo information is not correct because all the momenta are updated after the collision
step. If necessary, we suggest to derive these quantities from the output files.

\underline{Related commands:}

inlet/outlet interforce log output

\vspace{0.5cm}

\textbf{\underline{log command}}

\underline{Syntax:}

log(file)
\begin{itemize}
\item file=name of log file
\end{itemize}

\underline{Example:}

\&log("Log/log.poiseuille'');

\underline{Description:}

By default, the information printed on screen is also written in the file log.lb.
The command log allows to specify the name of the log file.

\underline{Related command:}

thermo

\vspace{0.5cm}

\textbf{\underline{output command}}

\underline{Syntax:}

output(file,$N_{\mathrm{f}}$)
\begin{itemize}
\item file=name of output file
\item $N_{\mathrm{f}}$=print output every this many timesteps
\end{itemize}

\underline{Example:}

\&output(``../Dump/output'',50);

\underline{Description:}

With this command it is specified the frequency with which information on the system state 
is printed in an output file. To the given file name is appended
the suffix ``.dat'' for dat files. The fields of every line of the output file
are defined as follows:\\
$x$ $y$ $\rho$ $v_{x}$ $v_{y}$\\
$x$ and $y$ are the coordinates of the lattice site, $\rho$ the
corresponding density, $v_{x}$ and $v_{y}$ the components of the
velocity vector, including also the contributions from the applied forces.
The first block of $(N_{x}+1)\times (N_{y}+1)$ lines defines the state
of the first frame; the second block of $(N_{x}+1)\times (N_{y}+1)$ lines defines the
state of the second frame; etc. This information refers to the state of the system
after the collision step (see Fig.~\ref{fig:algorithm}).

In the multicomponent case, an output file is written for every substance. The strings
``.fluid1.dat'' and ``.fluid2.dat'' are appended to the specified file name. In these
files there are four velocity components:\\
$x$ $y$ $\rho$ $v_{x}$ $v_{y}$ $v_{x}$ $v_{y}$\\ 
The first two velocity components refer to the
single substance. The last two values are the $x$ and $y$ components of the velocity of the 
overall fluid, Eq.~\ref{eq:u_solvent}.

If solute transport is enabled, an output file is written for the solute component. 
The string ``.solute.dat'' is appended to the specified file name. The fields of 
every line of the output file are defined as follows:\\
$x$ $y$ $C$ $b$\\
$x$ and $y$ are the coordinates of the lattice site and $C$ its solute concentration.
For solid sites, $b$ is the mass deposited on the reactive boundary; $b$ vanishes in
the fluid phase. This last quantity is omitted in the absence of surface reaction.
The data are organized into blocks as for the fluid phase. The information refers to 
the state of the system after the collision step and the cumulation of mass on the solid 
surface.

It is possible to write only one output file.

\vspace{0.5cm}

\textbf{\underline{write\_restart command}}

\underline{Syntax:}

write\_restart(file,$N_{\mathrm{f}}$)
\begin{itemize}
 \item file=name of restart file
\item $N_{\mathrm{f}}$=print restart file every this many timesteps
\end{itemize}

\underline{Example:}

\&write\_restart(``../Restart/restart.example'',100);

\underline{Description:}

This command defines the frequency with which a restart file is written.
The restart file is written at the end of the simulation if the number
of timesteps is a multiple of $N_{\mathrm{f}}$. To the given file name is appended
the string ``.timesteps'', where timestep is the timestep number at which
the restart file is written. The restart file has the format
of an input data file including all sections. The density value $-11$
is attributed to the sites belonging to the solid phase in the section
density, while all the velocity modes are set equal to $0$ in the
section momentum.

In the multicomponent case, a restart file is written for every substance. The
file name of the first component ends with the string ``.fluid1'' and that of
the second one ends with the string ``.fluid2''.

If solute transport is enabled, a restart file is written for the solute
component. The file name of the restart file ends with the string ``.solute''.

This command must be used before the iteration command.

\underline{Related commands:}

read\_data iteration

\vspace{0.5cm}

\textbf{\underline{iteration command}}

\underline{Syntax:}

iteration($N_{\mathrm{t}}$)
\begin{itemize}
\item $N_{\mathrm{t}}$=$\sharp$ of timesteps
\end{itemize}

\underline{Example:}

\&iteration(20000);

\underline{Description:}

This command fixes the number of iterations to be done. The iteration
command must follow any other command of the module. Without this
command the simulation is not performed.

\underline{Related command:}

write\_restart

\clearpage
\section{\label{sec:application}Applications}

\subsection{\label{sec:poiseuille}Poiseuille flow}

\begin{figure}[ht]
\begin{center}
\includegraphics[width=0.5\textwidth]{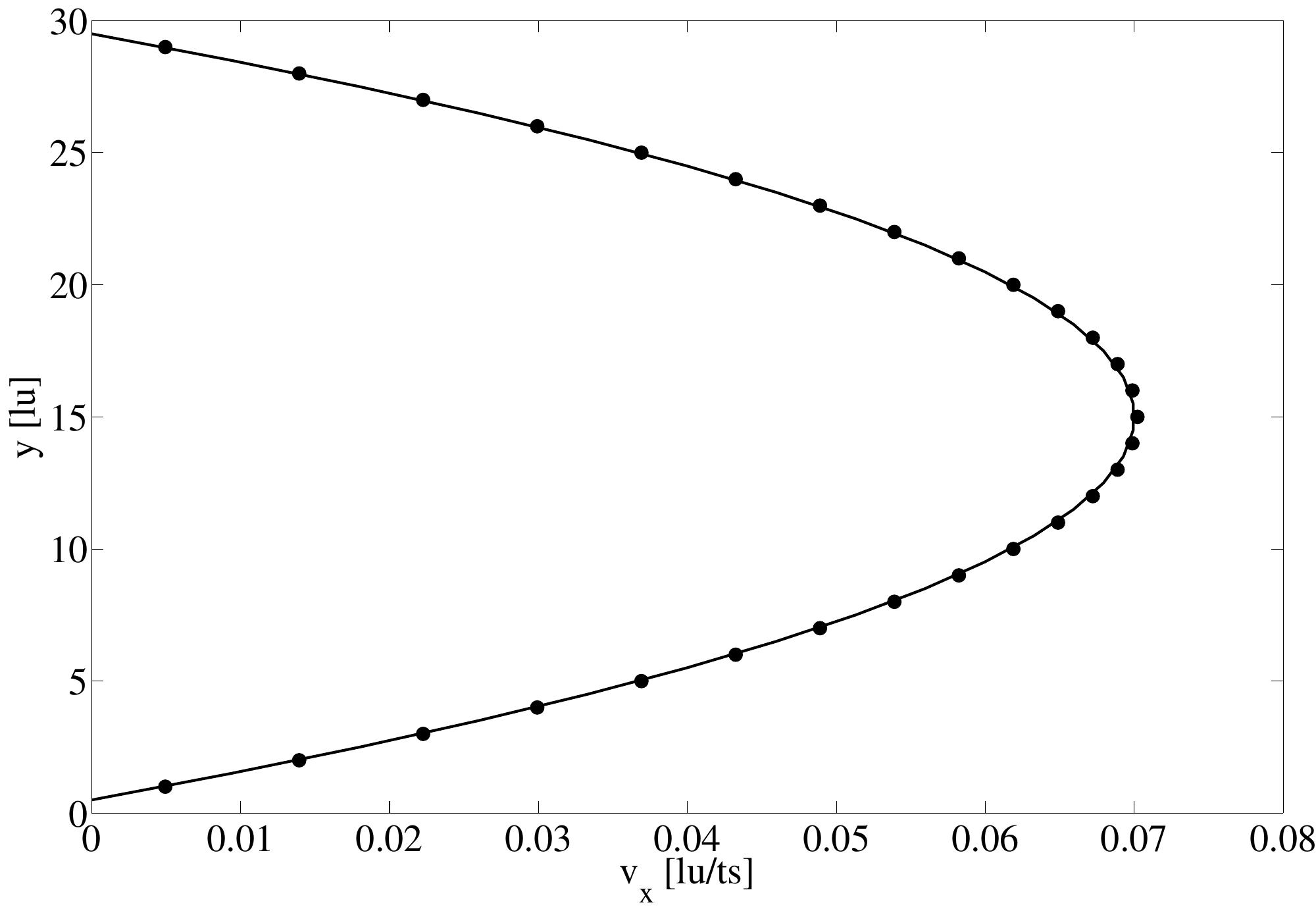}\hspace{0.5cm}
\end{center}
\caption{Velocity profile for Poiseuille flow. Points are obtained from
simulation. The solid line is the theoretical prediction, Eq.~\ref{eq:profile}.}
\label{fig:flow}
\end{figure}

A channel is obtained from the simulation domain with $N_{x}=60$ and
$N_{y}=30$ [lu] (d2q9 style). At $x=0$ and $N_{x}$ [lu] the boundaries are periodic
and at $y=0$ and $N_{y}$ [lu] the boundaries are solid, of type noslip. The channel
is filled with fluid: the initial populations of the rest and slow modes are set to
$4$ and those of the fast modes to $1$ [mu/lu$^{2}$]. A constant acceleration in the 
$\bm{e}_{1}$ direction is applied, of general form
\begin{equation*}
g=\frac{4}{3}\frac{u_{\mathrm{max}}}{(N_{y}-1)^{2}}(2\tau-1)\ .
\end{equation*}
We choose $u_{\mathrm{max}}=0.07$ [lu/ts] and $\tau=1$ [ts], leading to $g=1.11\cdot 10^{-4}$
[lu/ts$^{2}$]. In this way, the velocity component $v_{x}$ develops a parabolic profile in
the course of time. In the stationary regime, the theoretical velocity
profile is given by
\begin{equation}
v_{x}(y)=\frac{3g}{2\tau-1}\Big[\Big(\frac{N_{y}-1}{2}\Big)^2-\Big(y-\frac{N_{y}}{2}\Big)^{2}\Big]\ ,
\quad\text{with }y=\frac{1}{2},\dots,N_{y}-\frac{1}{2}\ .
\label{eq:profile}
\end{equation}
This behavior is usually referred to as Poiseuille flow. Figure \ref{fig:flow}
shows the velocity profile as obtained after $10'000$ timesteps.
\subsection{\label{sec:phase}Phase separation}

\begin{figure}[ht]
\begin{center}
\includegraphics[width=0.4\textwidth]{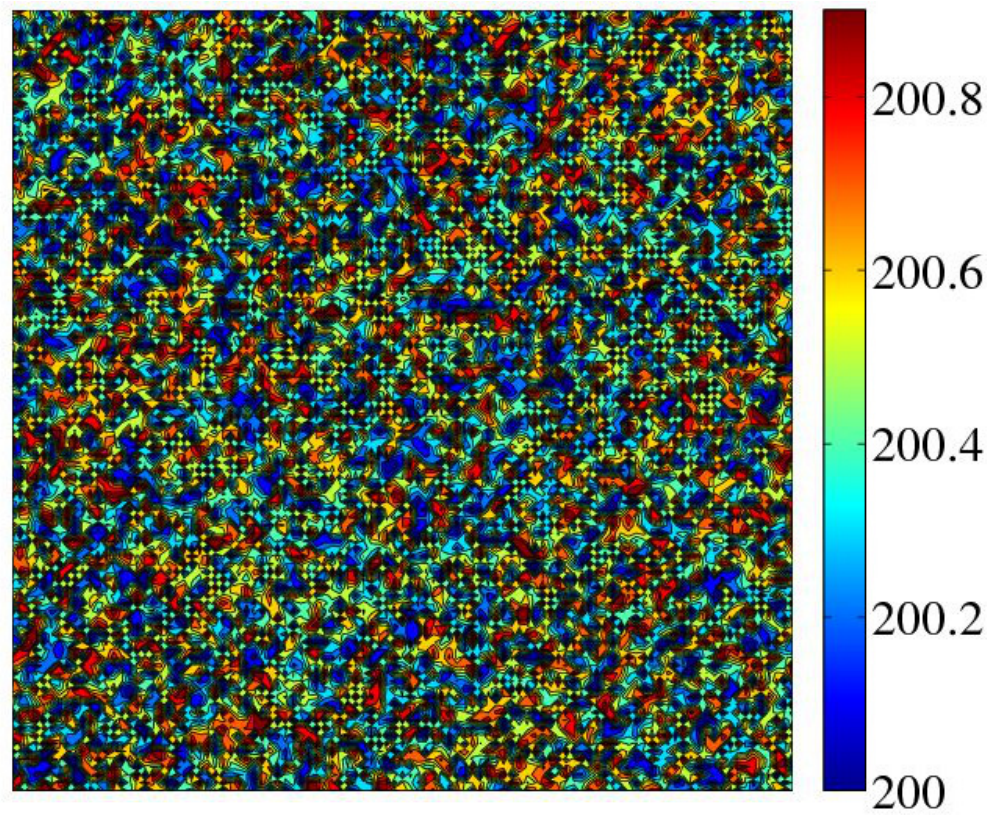}\hspace{0.5cm}
\includegraphics[width=0.387\textwidth]{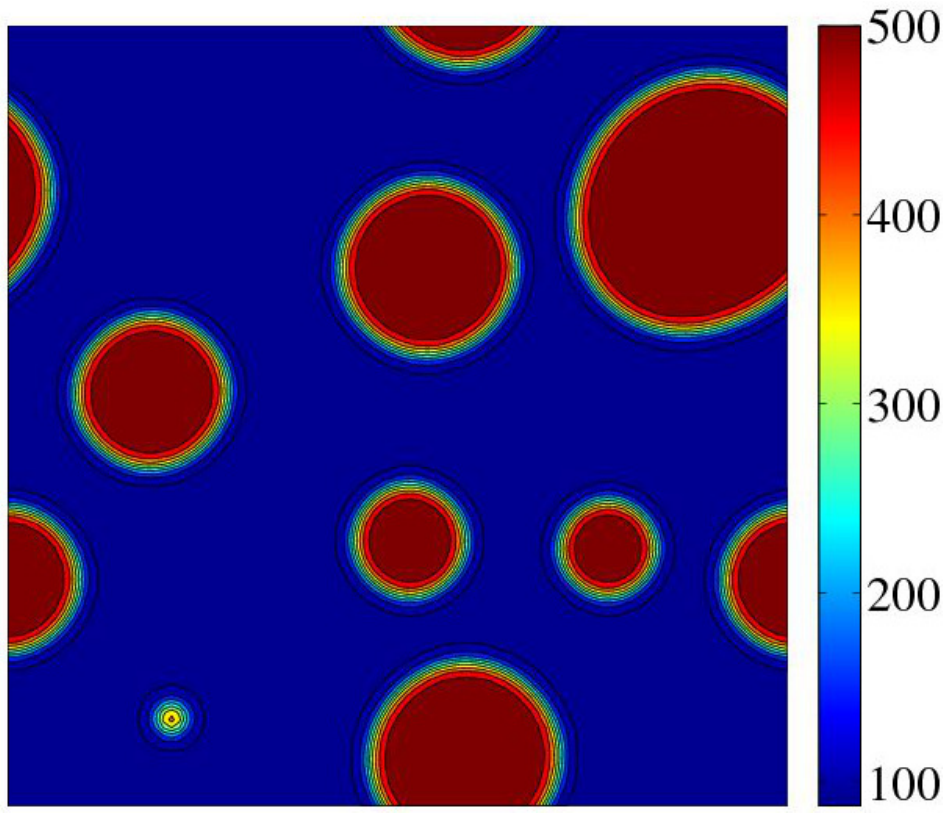}\\
\includegraphics[width=0.392\textwidth]{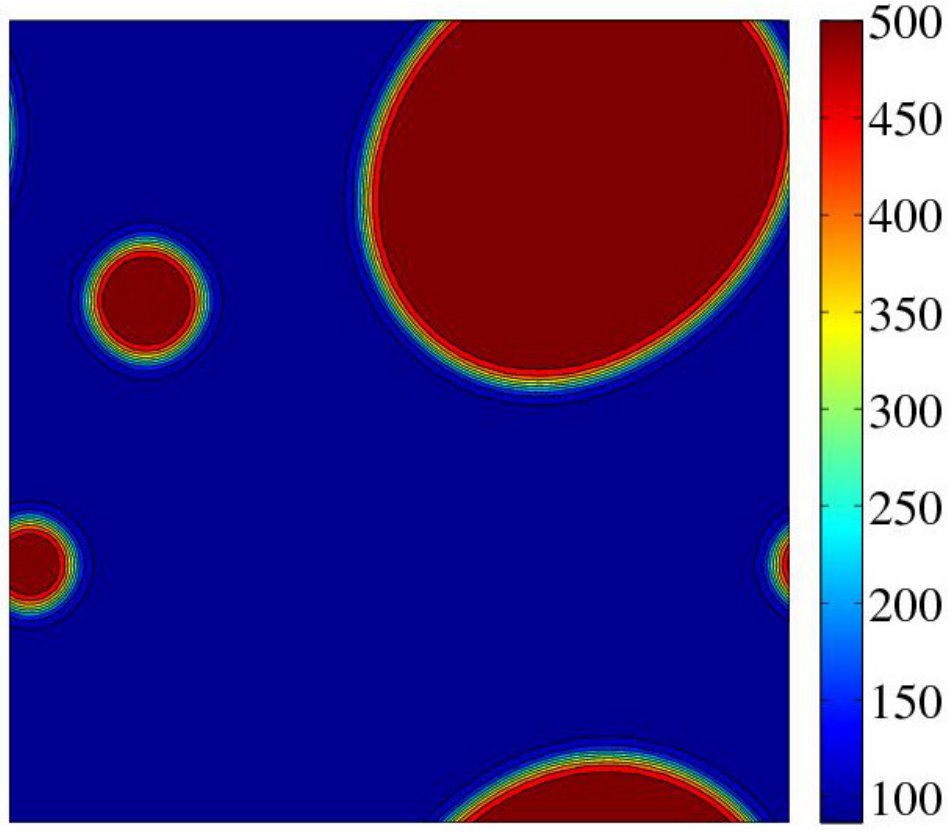}\hspace{0.5cm}
\includegraphics[width=0.389\textwidth]{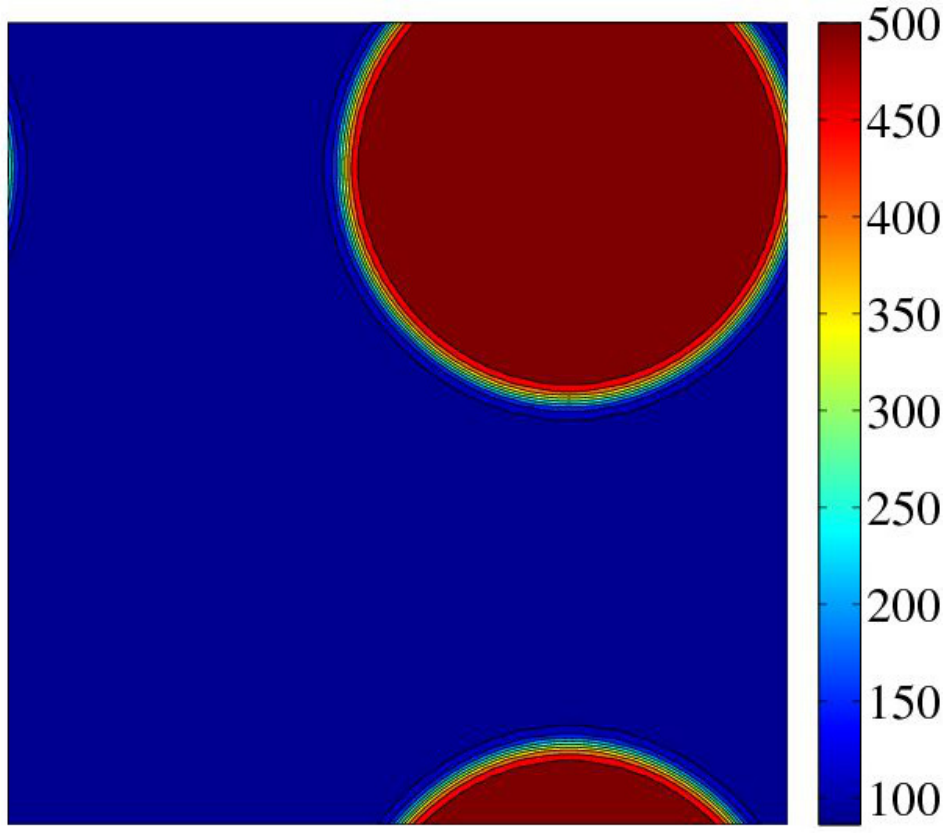}
\end{center}
\caption{Top: The state of the system at the beginning and after $2'000$
timesteps: many small droplets rapidly form. Bottom: State of the system
after $4'000$ and after $20'000$ timesteps. A single droplet forms for the
cohesive forces. Color map based on density.}
\label{fig:phase}
\end{figure}

Let us consider a lattice of style d2q9 with $N_{x}=N_{y}=100$ [lu]. The initial
density of every site is given by $d(x,y)=d_{0}+\delta d$ with $d_{0}=200$
[mu/lu$^{2}$] and $\delta d$ a random number in the interval $[0,1]$ (uniform distribution).
Cohesive forces are introduced via the interforce command with arguments $G=-120$ [mu$\cdot$lu/ts$^{2}$], 
$\psi_{0}=4$ and $\rho_{0}=200$ [mu/lu$^{2}$]. The system is let evolve for $20'000$ timesteps with
relaxation time $\tau=1$ [ts]. Figure \ref{fig:phase} shows four different frames
of the system dynamics. It clearly appears that the liquid phase,
higher density, separates from the vapor phase, lower density. For the
cohesive forces, small droplets merge together to form a single
droplet immersed in its own vapor phase.

\subsection{\label{sec:contactI}Contact angles I}

\begin{figure}[ht]
\begin{center}
\includegraphics[width=0.8\textwidth]{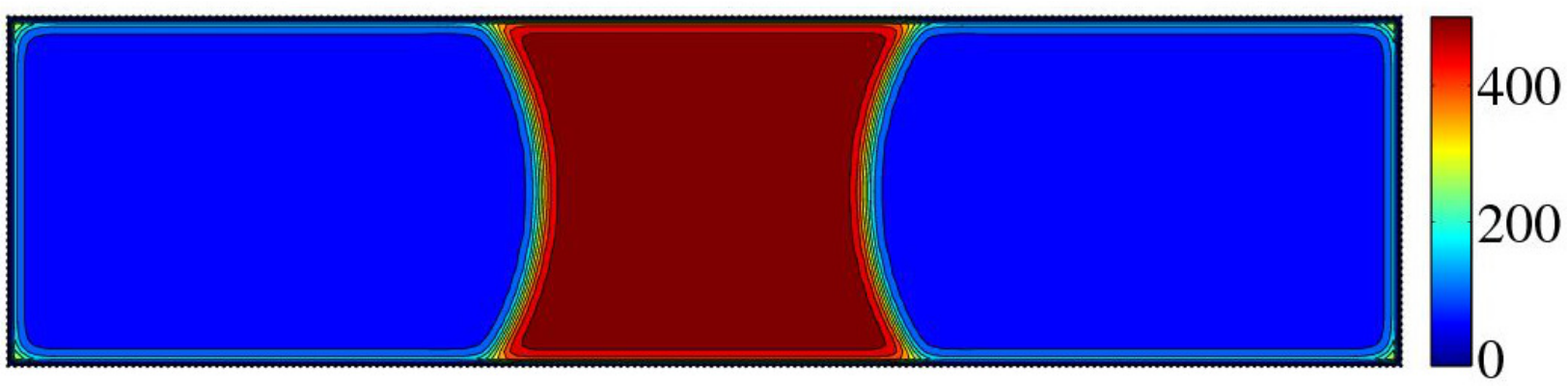}\\
\includegraphics[width=0.8\textwidth]{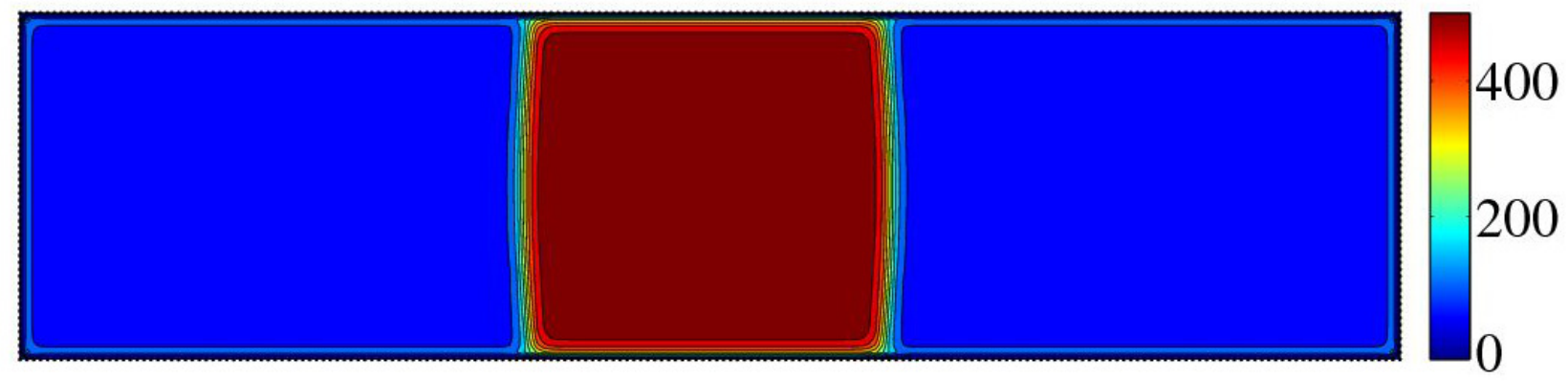}\\
\includegraphics[width=0.8\textwidth]{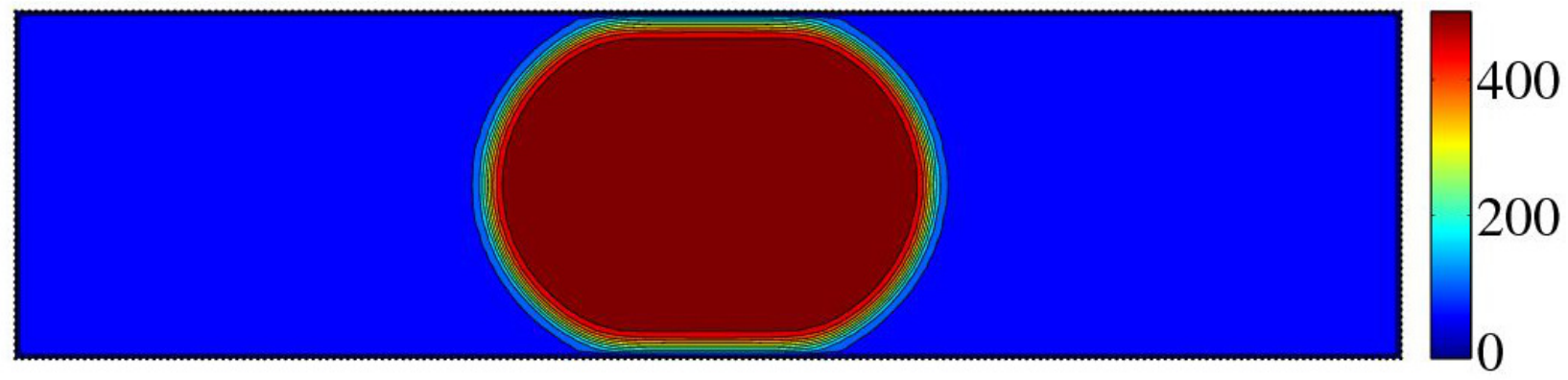}
\end{center}
\caption{Liquid phase confined in a rectangular box displays hydrophilic,
neutral and hydrophobic behavior, from Top to Bottom. The contact angle is
indeed $<90^{\circ}$, $\approx 90^{\circ}$ and $>90^{\circ}$, respectively. Color
map based on density. Black points represent solid boundaries.}
\label{fig:contact}
\end{figure}

We consider a fluid confined in a rectangular box of size
$(N_{x}+1)\times(N_{y}+1)$, with $N_{x}=200$ and $N_{y}=50$ [lu] (d2q9 style).
The boundaries of the simulation domain are solid, of type noslip, and
the initial density of the other lattice sites is initialized as in the
previous application. van der Waals forces are defined by the set of
parameters $G=-120$ [mu$\cdot$lu/ts$^{2}$], $\psi_{0}=4$ and $\rho_{0}=200$ [mu/lu$^{2}$].
Adhesive forces between the solid and liquid phases are introduced with the parameters 
$\psi_{0}=4$ and $\rho_{0}=200$ [mu/lu$^{2}$]. The interaction strength $G$ is varied in order to obtain
different contact angles \cite{gennes}. Figure \ref{fig:contact} shows the state
of the systems after $10'000$ timesteps; relaxation time $\tau=1$ [ts]. The
hydrophilic regime is reproduced with $G=-250$ [mu$\cdot$lu/ts$^{2}$]; $G=-187.16$ 
[mu$\cdot$lu/ts$^{2}$] leads to the neutral behavior; the hydrophobic regime is 
reproduced with $G=-70$ [mu$\cdot$lu/ts$^{2}$] \cite{book1}.

\subsection{\label{sec:contactII}Contact angles II}

\begin{figure}[ht]
\begin{center}
\includegraphics[width=0.8\textwidth]{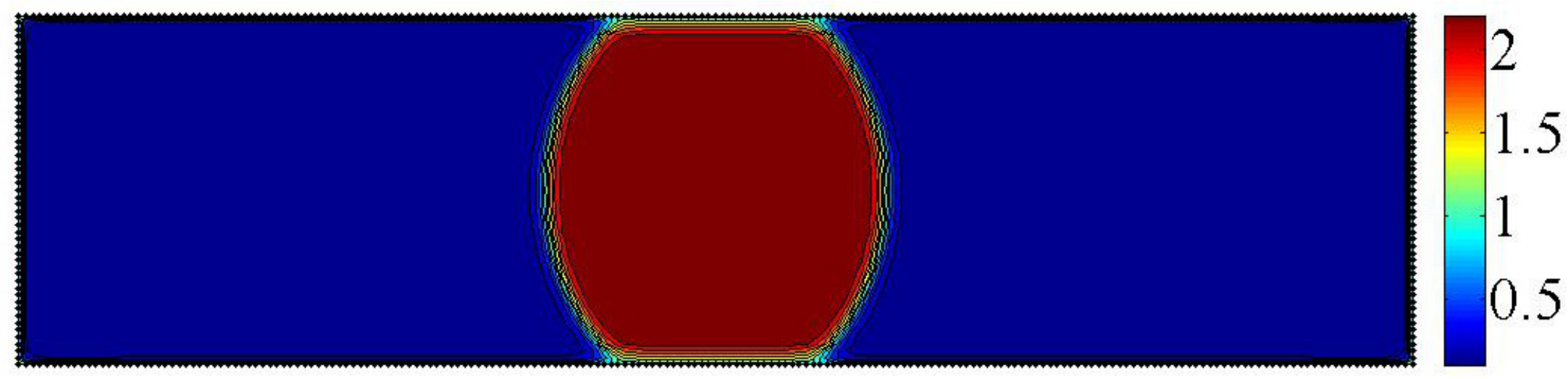}\\
\includegraphics[width=0.8\textwidth]{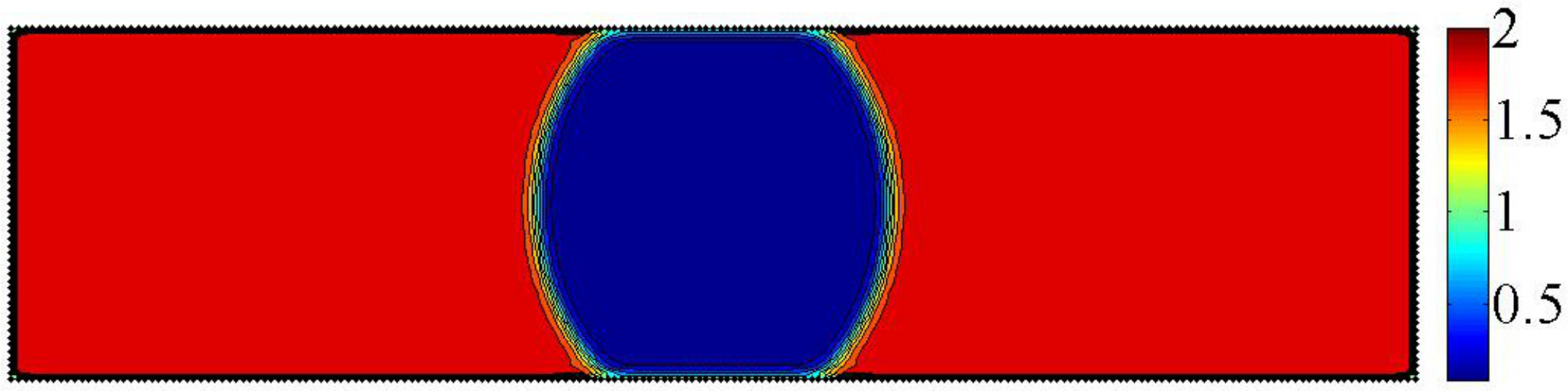}
\end{center}
\caption{
Final state for the solution of two immiscible substances.
Color map based on density; solid phase indicated by black points. Top: Representation
for the first fluid; Bottom: Representation for the second fluid. There remain two
distinct fluid components forming complementary contact angles with the
surface.}
\label{fig:fluids}
\end{figure}

Let us consider a rectangular domain with $N_{x}=200$ [lu] and $N_{y}=50$
[lu] with rigid boundaries of type noslip (d2q9 model). The simulation domain is
filled with two substances so that to be left with a single droplet of the first 
fluid in the middle. Inside a rectangle of base $51$ [lu] centered in the simulation
domain, the density for the first fluid is $f_{0}=2$ [mu/lu$^{2}$] and that of the second one $f_{0}=0.1$ 
[mu/lu$^{2}$]; elsewhere we choose $f_{0}=0.1$ [mu/lu$^{2}$] and $f_{0}=2$ [mu/lu$^{2}$] for the first 
and second component, respectively. Interparticle forces are introduced with parameter $G=0.9$ [lu/mu/ts$^{2}$] \cite{theta}. 
The interaction strength with the solid boundaries is $G_{1}=0.2$ [lu/ts$^{2}$] for the first fluid
and $G_{2}=-0.2$ [lu/ts$^{2}$] for the second fluid \cite{theta}. Figure \ref{fig:fluids}
shows the densities of the two substances after $20'000$ timesteps
(relaxation time $\tau=1$ [ts]). The fluid with lower density is interpreted
as dissolved in the other one. It appears that a single droplet forms in the middle 
with the first component exhibiting hydrophilic behavior while the second one hydrophobic.

\subsection{\label{sec:parallel}Parallelization}

\begin{figure}[ht]
\begin{center}
\includegraphics[width=0.5\textwidth]{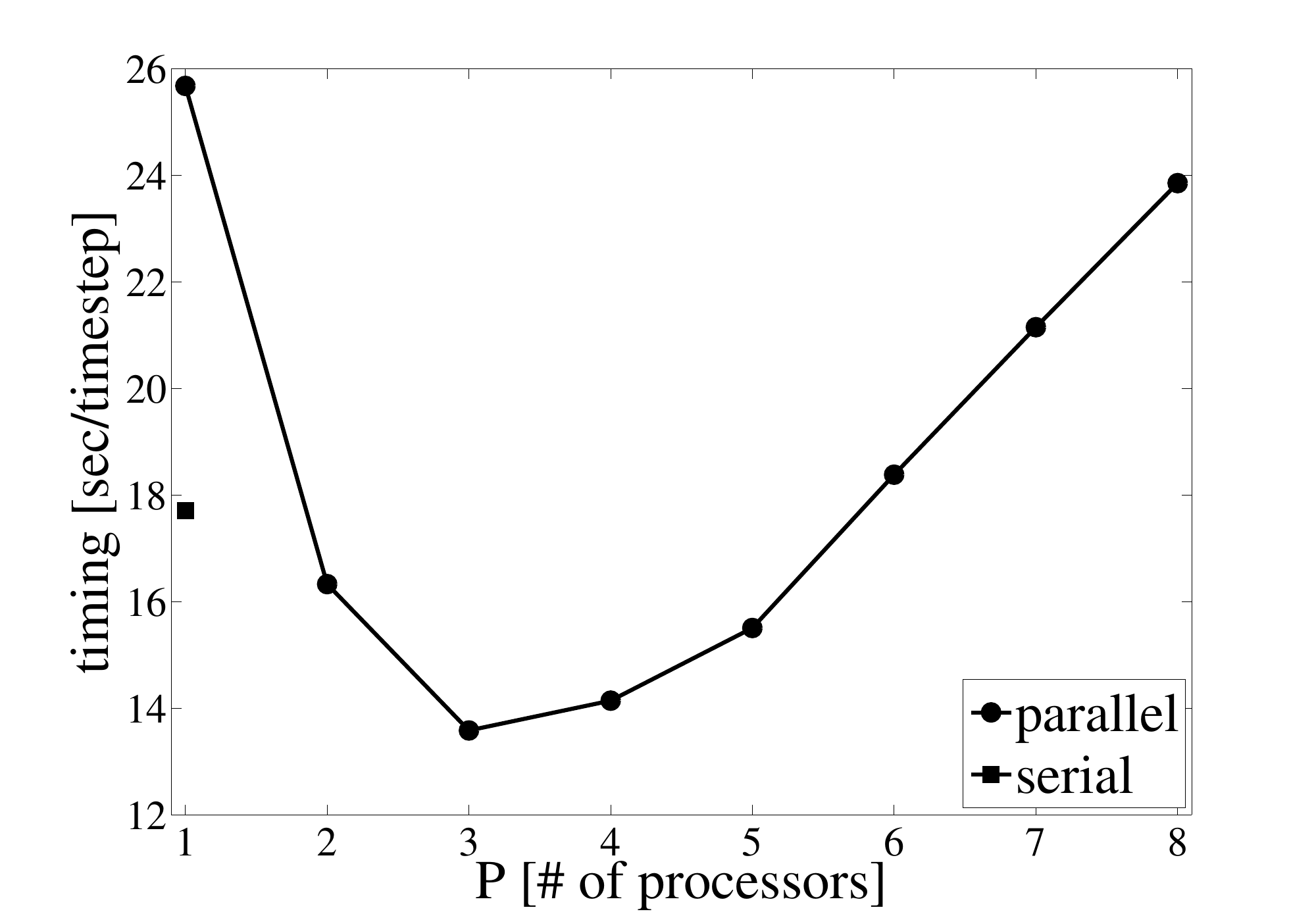}
\end{center}
\caption{
Timing for different processor grids. The parallelization for four or more processors is inefficient.}
\label{fig:cpu}
\end{figure}

\begin{figure}[hb]
\begin{center}
\includegraphics[width=0.5\textwidth]{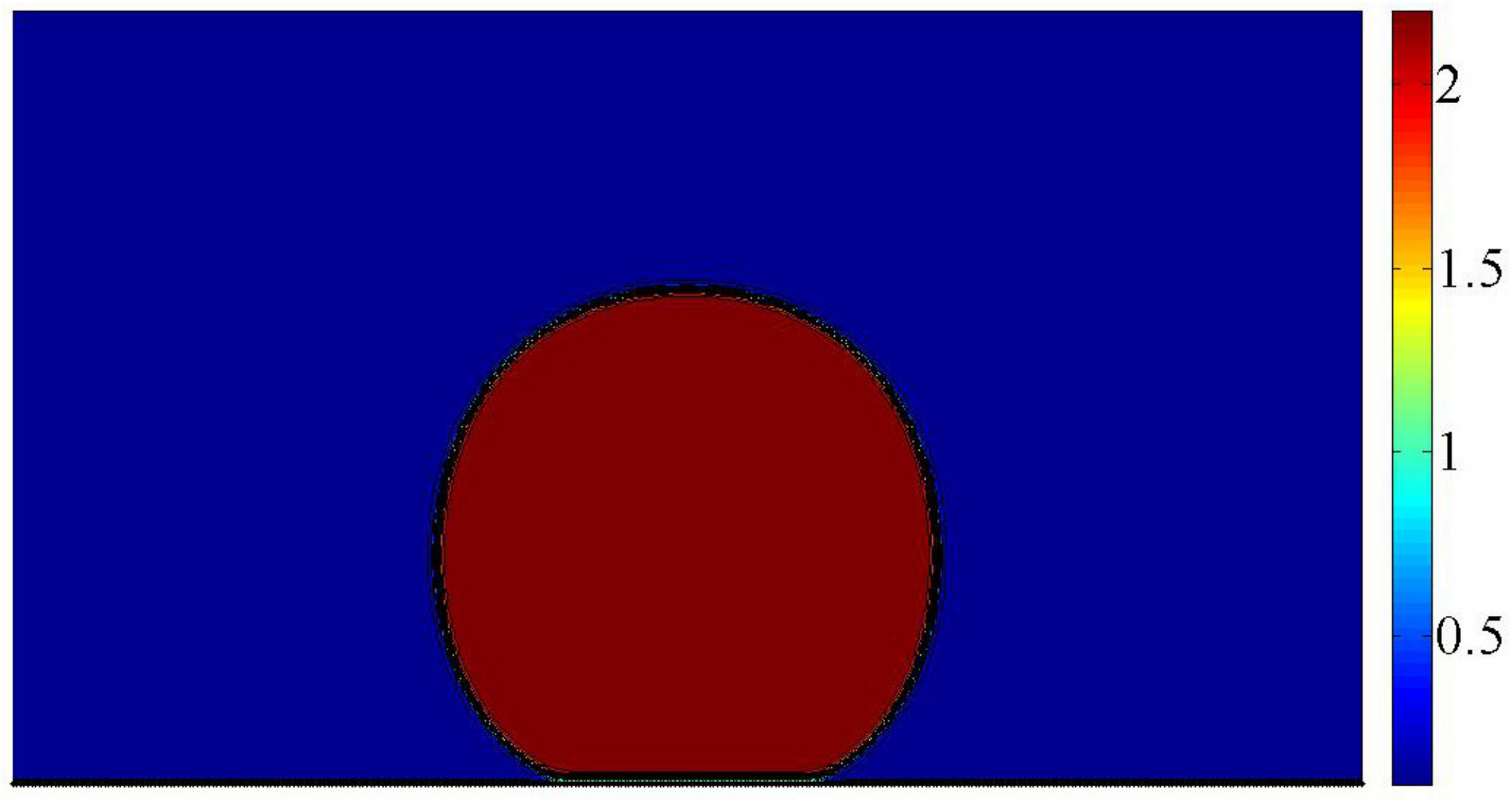}
\end{center}
\caption{
Fluid immersed in another one in contact with a solid boundary. 
A spherical droplet forms exhibiting hydrophobic behavior since for the 
contact angle holds $\theta>90^{\circ}$. Color map based on density; black
points represent solid nodes.}
\label{fig:droplet}
\end{figure}

Let us consider a two-component fluid confined in the simulation domain with $N_{x}=350$ [lu]
and $N_{y}=200$ [lu] (d2q9 model). The wall ylo is solid, of type noslip, and centered
above it we place a square of side length $129$ [lu]
with densities $f_{0}=2$ [mu/lu$^{2}$] for the first substance and $f_{0}=0.1$ [mu/lu$^{2}$] for the second one.
Elsewhere the densities are $f_{0}=0.1$ [mu/lu$^{2}$] and $f_{0}=2$ [mu/lu$^{2}$] for the first and second fluid, respectively.
These initial conditions guarantee the formation of a single droplet of the first component.
Interparticle forces are defined by $G=0.9$ [lu/mu/ts$^{2}$]; the interaction strengths for
adhesive forces are $G_{1}=0.3$ [lu/ts$^{2}$] and $G_{2}=-0.3$ [lu/ts$^{2}$] for the first and second
fluid component, respectively.
The wall xhi and xlo are periodic and the relaxation time is $\tau=1$ [ts]. The system is
let evolve for $500$ timesteps in the parallel mode with different number of processors;
the timings are shown in Fig.~\ref{fig:cpu}. We used a Desktop PC with operating system Windows 7 equipped
of an Intel processor of the family i7. Figure \ref{fig:droplet} shows the state of the system after $40'000$ 
timesteps.

\subsection{\label{sec:surface}Surface growth}

\begin{figure}[ht]
\begin{center}
\includegraphics[width=5cm]{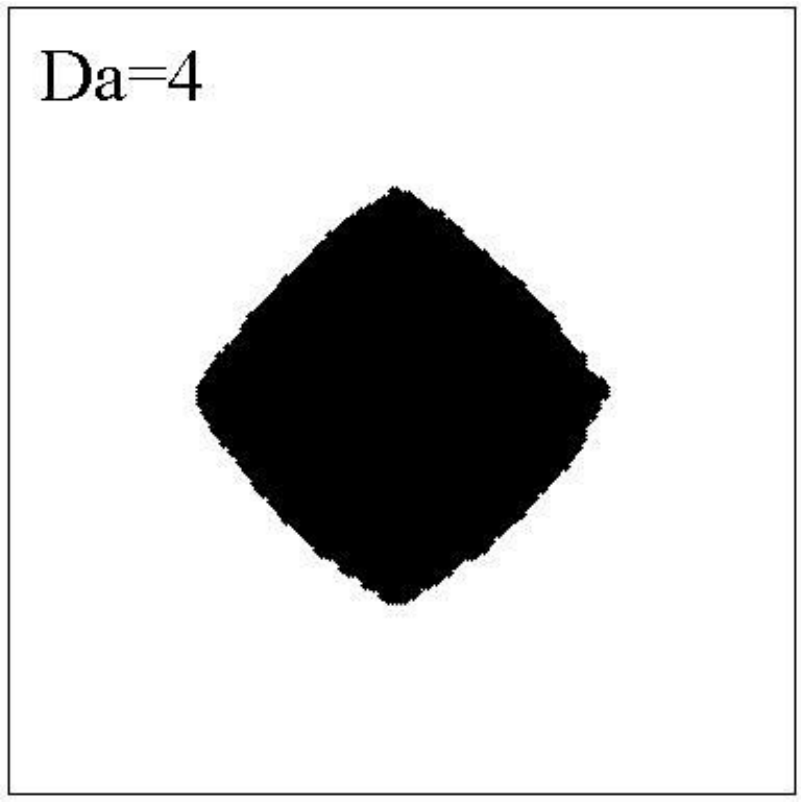}
\includegraphics[width=5cm]{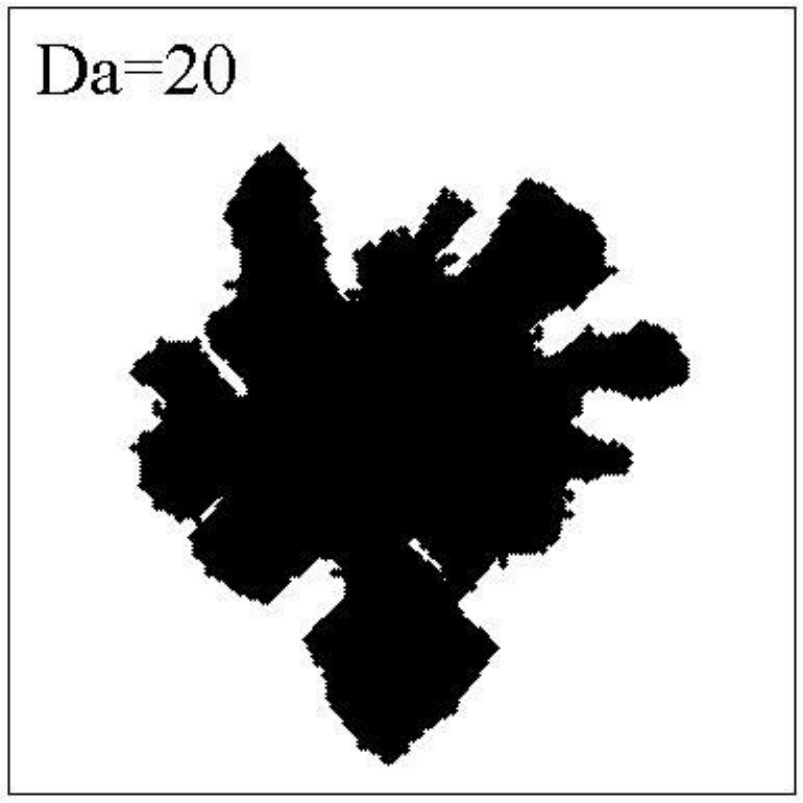}\\
\includegraphics[width=5cm]{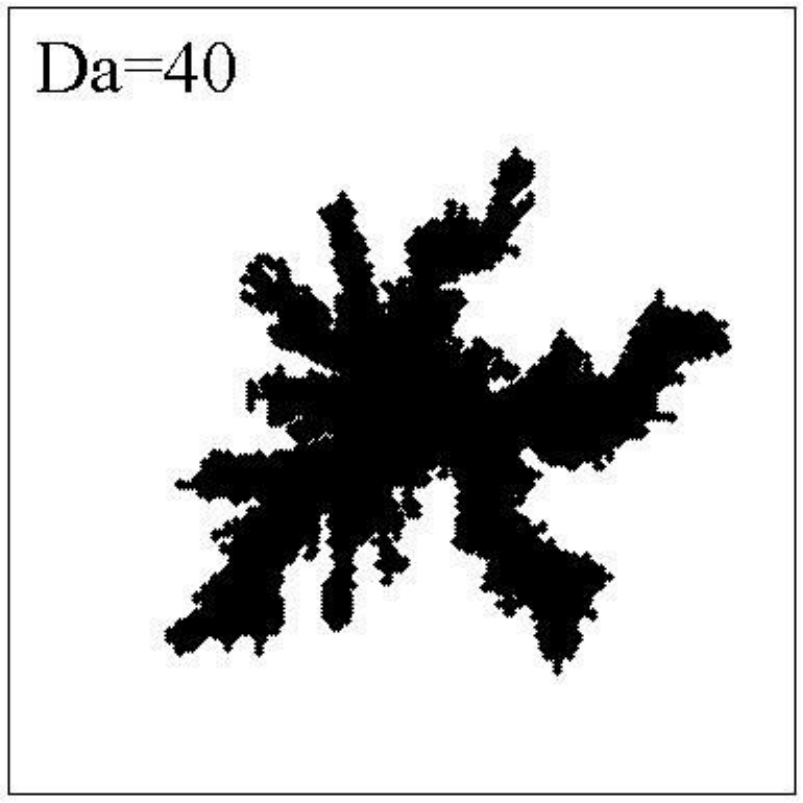}
\includegraphics[width=5cm]{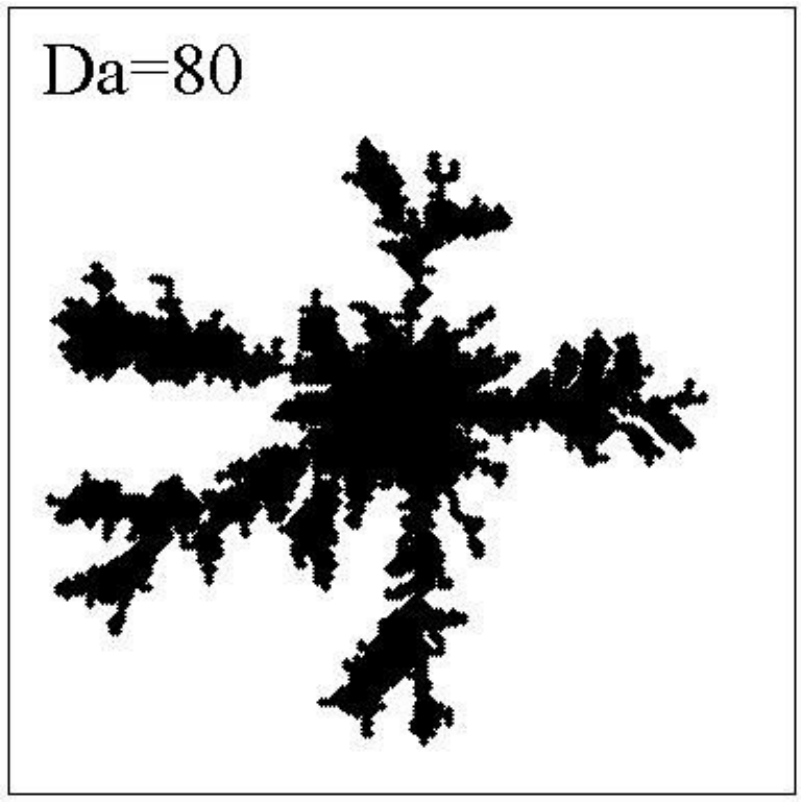}
\end{center}
\caption{
Surface growth in the purely diffusive regime from a small seed. Final configuration for systems with different 
Damkohler number $Da$. As the Damkohler number increases the structures become more dendritic or,
equivalently, less compact.}
\label{fig:surface}
\end{figure}

The simulation domain is $N_{x}=200$ [lu] long and $N_{y}=200$ [lu] wide with periodic boundary conditions (d2q9 model).
A solid obstacle of square shape with side length $11$ [lu] is placed in the center. The collisions with the solid boundaries 
are treated according to the bounce-back rule corresponding to the noslip style. The rest of the simulation domain is filled uniformly 
with a single component fluid and solute. The initial density of the fluid is defined by $f_{0}=24$ [mu/lu$^{2}$]. The initial concentration 
of the solute is $C_{0}=1.5$ [mu/lu$^{2}$]. The parameters controlling the surface reaction are chosen as follows. The saturated 
concentration is set to $C_{\mathrm{s}}=0.5$ [mu/lu$^{2}$] and the intial mass on the solid surface is $b_{0}=0.5$ [mu/lu$^{2}$]. The surface 
grows whenever the cumulated mass on the solid surface exceeds the threshold value of $b_{\mathrm{max}}=1$ [mu/lu$^{2}$]. For the reaction-rate 
constant $k_{\mathrm{r}}$, different values are considered. The dynamics amounts to $20'000$ timesteps with relaxation times $\tau=1$ [ts] 
for fluid flow and $\tau_{\mathrm{s}}=1$ [ts] for solute transport. The properties of the growing surface depend on the relative saturation
$\psi=C_{0}/C_{\mathrm{s}}$ and the Damkohler number $Da=k_{\mathrm{r}}L/D$. $L$ is a characteristic length of the system and $D$ the diffusion 
coefficient. Figure \ref{fig:surface} shows the resulting final states for different Damkohler numbers.
\clearpage
\section{\label{sec:script}Scripts}

\subsection{\label{sec:scriptflow}Poiseuille flow}

The data of Sec.~\ref{sec:poiseuille} were generated with the following Perl program.
\begin{verbatim}
#!/usr/bin/perl
#======================================================================

use lib "D:\\LB_data\\Module";      #perl -e 'print join "\n", @INC'

use strict;
use lb2d qw(
lattice processors read_data fluids solute inlet outlet inlet_momentum
outlet_momentum inlet_solute outlet_solute boundary_style obstacle 
surface_reaction position0 momentum0 concentration0

aveforce aveacceleration interforce adhesiveforce intersolute

thermo log output write_restart

iteration);
#======================================================================
{
#======================================================================

    my $Nx=60;
    my $Ny=30;
    my $Nt=10000;

    &lattice("d2q9",$Nx,$Ny,1);	 

    &boundary_style("noslip");
    &obstacle("rectangle",0,$Nx,0,0);
    &obstacle("rectangle",0,$Nx,$Ny,$Ny);

    &position0("rectangle",0,$Nx,1,$Ny-1); 

    &momentum0("e0",4);
    &momentum0("e1",4); 
    &momentum0("e2",4);
    &momentum0("e3",4);
    &momentum0("e4",4);
    &momentum0("e5",1);
    &momentum0("e6",1);
    &momentum0("e7",1);
    &momentum0("e8",1);

    &aveacceleration("e1",1.11*10**-4);

    &thermo(100);
    &log("log\.poiseuille");
    &output("v_profile",int($Nt/20));

    &write_restart("restart.v_profile",$Nt);

    &iteration($Nt);

#======================================================================
}
\end{verbatim}

The command \verb1use lib "D:\\LB\_data\\Module"1 specifies the location of the module.
If the command under commentary is typed in the command line, it is possible
to have a list of the directories that Perl scans by default. If the module
is in one of these directories, this line is not necessary. With the command
\verb1use lb2d qw(...)1, the listed functions of the module lb2d.pm are loaded.
This command is of initialization and must appear in any program.

All the commands of the module are used in a straightforward way. Note that
operations and variables can be used as arguments of the functions. As an example,
in the command output, with \verb1int($Nt/20)1 it is specified to write in the file
v\_profile $20$ evenly spaced frames.

The raw data can be analyzed with the following Matlab program.
\begin{verbatim}
clear;
clf;
%==============================================

frame=21;
Nx=60;
Ny=30;
tau=1;
g=1.11*10^-4;

nu=(tau-0.5)/3;

nodes=(Nx+1)*(Ny+1);

%==============================================

load D:\\LB_data\Poiseuille\v_profile.dat;

x=[v_profile(:,1)];
y=[v_profile(:,2)];
d=[v_profile(:,3)];
ux=[v_profile(:,4)];
uy=[v_profile(:,5)];

%==============================================

X=0:1:Nx;
Y=0:1:Ny;

[A,B]=meshgrid(X,Y);
[sx,sy]=meshgrid(0:Nx,0:Ny);

for i=1:frame
   from=1+(i-1)*nodes;
   to=nodes+(i-1)*nodes;
   k=0;
   q=0;
   for j=from:to
      if d(j)==-11
         q=q+1; 
         ox(q,i)=x(j);
         oy(q,i)=y(j);   
      end
      k=k+1;
      u(k,i)=x(j);
      v(k,i)=y(j);
      vx(k,i)=ux(j);
      vy(k,i)=uy(j);
      z1(y(j)+1,x(j)+1,i)=d(j);
      z2(y(j)+1,x(j)+1,i)=ux(j); 
      z3(y(j)+1,x(j)+1,i)=uy(j);
      z4(y(j)+1,x(j)+1,i)=sqrt(ux(j)^2+uy(j)^2);
   end
end

%==============================================

figure(1);
hold on;
box on;
axis equal;
for i=1:frame
   str=sprintf('frame: %d',i-1);
   title(str);
   colorbar;
   cla;
   colormap(jet);
   contourf(X,Y,z2(:,:,i));
   quiver(u(:,i),v(:,i),vx(:,i),vy(:,i),'w');
   %streamline(A,B,z2(:,:,i),z3(:,:,i),sx,sy);
   plot(ox(:,i),oy(:,i),'k.');
   axis([0 Nx 0 Ny -50 50]);
   saveas(1,strcat('v_profile',num2str(i-1)),'jpg');
   pause(0.1);
end

ly=-(Ny-1)/2:1:(Ny-1)/2;
vv=((Ny-1)/2)^2-ly.^2;
vv=g*vv./(2*nu);

figure(2);
hold on;
box on;
plot(vx(2:Ny,frame),1:1:Ny-1,'ko');
plot(vv,ly+Ny/2,'k-');
\end{verbatim}

In the first block are entered the general settings of the Lattice
Boltzmann simulation. In the second block, the data are loaded.
In the third block, the data are read and organized into matrices
for the usual quantities. The last entry of these matrices always refers
to the frame number. Finally, the data are displayed frame after
frame with the instructions of the last block. The color map is based
on the $x$ component of velocity (contourf command) and the vector field is also
given by the velocity vector (quiver command). By enabling the streamline command,
streamlines derived from the velocity field are drawn. With the command saveas, a
figure in the format jpg is created for every frame. The second figure displays the 
velocity profile
(see Fig.~\ref{fig:flow}).
\subsection{Phase separation}

The following Perl commands generate the data presented in Sec.~\ref{sec:phase}.
\begin{verbatim}	
    my $Nx=100;
    my $Ny=100;
    my $Nt=20000;

    &lattice("d2q9",$Nx,$Ny,1);

    &read_data("phase_data","equilibrium");

    &interforce(-120,4,200);

    &log("log\.phase");
    &thermo(100);
    &output("phase",int($Nt/20));

    &write_restart("phase",$Nt);

    &iteration($Nt);
\end{verbatim}
The initialization is omitted since already discussed for the first example. In this case, the 
initial state of the system is imported from the input data file phase\_data, read with the command 
read\_data. At the beginning, the distribution functions are replaced with their equilibrium
values because of the additional argument equilibrium.

The raw data are analyzed with a Matlab program similar to that given
for Poiseuille flow.

\subsection{Contact angles I}

The Perl commands specific to the case of hydrophilic behavior are
given below (see Sec.~\ref{sec:contactI}).
\begin{verbatim}
    my $Nx=200;
    my $Ny=50;
    my $Nt=10000;

    &lattice("d2q9",$Nx,$Ny,1);	 

    &read_data("box_data","equilibrium");

    &boundary_style("noslip");
    &obstacle("rectangle",0,0,0,$Ny);
    &obstacle("rectangle",$Nx,$Nx,0,$Ny);
    &obstacle("rectangle",0,$Nx,0,0);
    &obstacle("rectangle",0,$Nx,$Ny,$Ny);

    &interforce(-120,4,200);
    &adhesiveforce(-250,4,200);

    &log("log\.box_philic");
    &thermo(100);
    &output("box_philic",int($Nt/20));

    &write_restart("restart.box_philic",$Nt);

    &iteration($Nt);
\end{verbatim}
Note that the obstacle commands override the definitions of density
and momentum with the command read\_data.

The raw data can be analyzed after small changes to the Matlab program
for Poiseuille flow (see Sec.~\ref{sec:scriptflow}).
\subsection{Contact angles II}

The data of the application of Sec.~\ref{sec:contactII} are generated with
the following Perl program.
\begin{verbatim}
#!/usr/bin/perl
#======================================================================

use lib "D:\\LB_data\\Module";      #perl -e 'print join "\n", @INC'

use strict;
use lb2d_par qw(
lattice processors read_data fluids solute inlet outlet inlet_momentum
outlet_momentum inlet_solute outlet_solute boundary_style obstacle 
surface_reaction position0 momentum0 concentration0

aveforce aveacceleration interforce adhesiveforce intersolute

thermo log output write_restart

iteration);
#======================================================================
{
#======================================================================

    my $Nx=200;
    my $Ny=50;
    my $Nt=20000;

    my $fluid1=1;
    my $fluid2=2;
    my $tau1=1;
    my $tau2=1;

    &lattice("d2q9",$Nx,$Ny,1);	 

    &processors(2,1);	

    &fluids($tau1,$tau2);

    &boundary_style("noslip");
    &obstacle("rectangle",0,$Nx,0,0);
    &obstacle("rectangle",0,$Nx,$Ny,$Ny);
    &obstacle("rectangle",0,0,0,$Ny);
    &obstacle("rectangle",$Nx,$Nx,0,$Ny);

    &position0("rectangle",1,$Nx/2-26,1,$Ny-1); 
    &momentum0("e0",0.1,$fluid1);
    &momentum0("e0",2,$fluid2);

    &position0("rectangle",$Nx/2-25,$Nx/2+25,1,$Ny-1); 
    &momentum0("e0",2,$fluid1);
    &momentum0("e0",0.1,$fluid2);

    &position0("rectangle",$Nx/2+26,$Nx,1,$Ny-1); 
    &momentum0("e0",0.1,$fluid1);
    &momentum0("e0",2,$fluid2);

    &interforce(0.9);
    &adhesiveforce(0.2,$fluid1);
    &adhesiveforce(-0.2,$fluid2);

    &log("log\.theta_fluids");
    &thermo(100);
    &output("theta_fluids",int($Nt/20));

    &write_restart("restart.theta_fluids",$Nt);

    &iteration($Nt);

#======================================================================
}
\end{verbatim}
The command \verb1use lb2d_par qw(...)1 loads the parallel version of the module.
The list of functions is the same as for the serial version.
The simulation is run on two CPUs (processors command). It should be noted that
the commands momentum0 and adhesiveforce have one extra argument specifying
the fluid component since two substances are present (fluids command). 
On a cluster with a queue system, it is necessary to submit the job with a shell
script similar to the following.
\begin{verbatim}
#!/bin/sh
#$ -N fluids
#$ -S /bin/sh
#$ -cwd
#$ -j y
#$ -q long.q
#$ -pe smp 2
#$ -l h_rt=24:00:00

echo "Starting job fluids at " `date`
mpirun -np $NSLOTS perl < theta_fluids.pl
echo "Ending job fluids at " `date`
\end{verbatim}
The second line attributes the name fluids to the job; the third line
specifies the shell interpreter; the current directory becomes the working
directory with line $4$; with the next line, STDERR is redirected to STDOUT;
the sixth line selects the queue long.q; at line $7$ the user requests
two CPUs on the same node and the variable \$NSLOTS is set to this value; at
line $8$ it is specified the walltime. With the command mpirun, the Perl program
is executed in parallel on \$NSLOTS cores. The echo commands print in the STDOUT
file the local time at the beginning and at the end of the simulation.

The data can be visualized and analyzed with a Matlab program similar to that
of Sec.~\ref{sec:scriptflow}.

\subsection{\label{sec:bench}Parallelization}

The data of Fig.~\ref{fig:droplet} in Sec.~\ref{sec:parallel} were generated with the following
Perl program.
\begin{verbatim}
#!/usr/bin/perl
#======================================================================

use lib "D:\\LB_data\\Module";      #perl -e 'print join "\n", @INC'

use strict;
use lb2d_par qw(
lattice processors read_data fluids solute inlet outlet inlet_momentum
outlet_momentum inlet_solute outlet_solute boundary_style obstacle 
surface_reaction position0 momentum0 concentration0

aveforce aveacceleration interforce adhesiveforce intersolute

thermo log output write_restart

iteration);
#======================================================================
{
#======================================================================

    my $Nx=350;
    my $Ny=200;
    my $Nt=20000;

    my $L=64;

    my $fluid1=1;
    my $fluid2=2;
    my $tau1=1;
    my $tau2=1;

    &lattice("d2q9",$Nx,$Ny,1);	 

    &processors(3,1);	

    &fluids($tau1,$tau2);

    &boundary_style("noslip");
    &obstacle("rectangle",0,$Nx,0,0);

    &position0("rectangle",0,$Nx,2*$L+1,$Ny); 
    &momentum0("e0",0.1,$fluid1);
    &momentum0("e0",2,$fluid2);

    &position0("rectangle",0,$Nx/2-$L-1,0,2*$L); 
    &momentum0("e0",0.1,$fluid1);
    &momentum0("e0",2,$fluid2);

    &position0("rectangle",$Nx/2+$L+1,$Nx,0,2*$L); 
    &momentum0("e0",0.1,$fluid1);
    &momentum0("e0",2,$fluid2);

    &position0("rectangle",$Nx/2-$L,$Nx/2+$L,0,2*$L); 
    &momentum0("e0",2,$fluid1);
    &momentum0("e0",0.1,$fluid2);

    &interforce(0.9);
    &adhesiveforce(0.3,$fluid1);
    &adhesiveforce(-0.3,$fluid2);

    &log("log\.droplet_fluids");
    &thermo(100);
    &output("droplet_fluids",$Nt);

    &write_restart("restart.droplet_fluids",$Nt);

    &iteration($Nt);

#======================================================================
}
\end{verbatim}
This simulation can be continued for other $20'000$ timesteps with the
following commands.
\begin{verbatim}
    my $Nx=350;
    my $Ny=200;
    my $Nt=20000;

    &lattice("d2q9",$Nx,$Ny,1);	 

    &read_data("restart.droplet_fluids.20000","fluids");	

    &log("log\.droplet2_fluids");
    &thermo(100);
    &output("droplet2_fluids",$Nt);

    &write_restart("restart.droplet2_fluids",$Nt);

    &iteration($Nt);
\end{verbatim}
The data were visualized with a Matlab program similar to that discussed in Sec.~\ref{sec:scriptflow}.

\subsection{Surface growth}

The simulations presented in Sec.~\ref{sec:surface} were carried out using the following
Perl commands.
\begin{verbatim}
#======================================================================
#
# label		
# 0 --> Da=4    kr=5*10**-3
# 1 --> Da=20   kr=0.025
# 2 --> Da=40   kr=0.05
# 3 --> Da=80   kr=0.1
#
#======================================================================

    my $label=0;		

    my $Nx=200;
    my $Ny=200;
    my $Nt=20000;

    my $tau=1;

    my $taus=1;
    my $mass0=0.5;
    my $mass1=2*$mass0;
    my $kr=5*10**-3;
    my $Cs=0.5;
    my $C0=1.5;

    &lattice("d2q9",$Nx,$Ny,$tau);

    &processors(2,1);

    &solute($taus);

    &surface_reaction($kr,$Cs,"growth",$mass0,$mass1);

    &boundary_style("noslip");
    &obstacle("rectangle",100-5,100+5,100-5,100+5);

    &position0("rectangle",0,$Nx,0,$Ny);
    &momentum0("e0",24);
    &concentration0($C0);

    &thermo(1000);
    &log("log\.growth".$label);
    &output("growth".$label,int($Nt/20));
    &write_restart("growth".$label,$Nt);

    &iteration($Nt);
\end{verbatim}
The module is loaded as for the previous examples. The commands solute and concentration0 initialize solute 
transport. The parameters controlling the surface reaction are entered via the command surface\_reaction. 

The data were displayed with a Matlab program similar to that discussed in Sec.~\ref{sec:scriptflow}.

\clearpage
\appendix

\section{Formulas for inlet/outlet openings in d2q9 model}\label{ap:inout}

For the inlets on the other walls, when the velocity is prescribed, we use
\begin{eqnarray*}
&&\rho=\frac{1}{1-v_{x}}[f_{0}+f_{2}+f_{4}+2(f_{1}+f_{5}+f_{8})]\ ,\\
&&f_{3}=f_{1}+\frac{2}{3}\rho v_{x}\ ,\\
&&f_{6}=f_{8}+\frac{1}{2}(f_{4}-f_{2})+\frac{1}{6}\rho v_{x}\ ,\\
&&f_{7}=f_{5}-\frac{1}{2}(f_{4}-f_{2})+\frac{1}{6}\rho v_{x}\quad\quad\text{for xhi;}
\end{eqnarray*}
\begin{eqnarray*}
&&\rho=\frac{1}{1-v_{y}}[f_{0}+f_{1}+f_{3}+2(f_{2}+f_{5}+f_{6})]\ ,\\
&&f_{4}=f_{2}+\frac{2}{3}\rho v_{y}\ ,\\
&&f_{7}=f_{5}+\frac{1}{2}(f_{1}-f_{3})+\frac{1}{6}\rho v_{y}\ ,\\
&&f_{8}=f_{6}-\frac{1}{2}(f_{1}-f_{3})+\frac{1}{6}\rho v_{y}\quad\quad\text{for yhi;}
\end{eqnarray*}
\begin{eqnarray*}
&&\rho=\frac{1}{1-v_{y}}[f_{0}+f_{1}+f_{3}+2(f_{4}+f_{7}+f_{8})]\ ,\\
&&f_{2}=f_{4}+\frac{2}{3}\rho v_{y}\ ,\\
&&f_{5}=f_{7}+\frac{1}{2}(f_{3}-f_{1})+\frac{1}{6}\rho v_{y}\ ,\\
&&f_{6}=f_{8}-\frac{1}{2}(f_{3}-f_{1})+\frac{1}{6}\rho v_{y}\quad\quad\text{for ylo.}
\end{eqnarray*}
In the case of outlet, the formulas read
\begin{eqnarray*}
&&\rho=\frac{1}{1+v_{x}}[f_{0}+f_{2}+f_{4}+2(f_{1}+f_{5}+f_{8})]\ ,\\
&&f_{3}=f_{1}-\frac{2}{3}\rho v_{x}\ ,\\
&&f_{6}=f_{8}+\frac{1}{2}(f_{4}-f_{2})-\frac{1}{6}\rho v_{x}\ ,\\
&&f_{7}=f_{5}-\frac{1}{2}(f_{4}-f_{2})-\frac{1}{6}\rho v_{x}\quad\quad\text{for xhi;}
\end{eqnarray*}
\begin{eqnarray*}
&&\rho=\frac{1}{1+v_{x}}[f_{0}+f_{2}+f_{4}+2(f_{3}+f_{6}+f_{7})]\ ,\\
&&f_{1}=f_{3}-\frac{2}{3}\rho v_{x}\ ,\\
&&f_{5}=f_{7}-\frac{1}{2}(f_{2}-f_{4})-\frac{1}{6}\rho v_{x}\ ,\\
&&f_{8}=f_{6}+\frac{1}{2}(f_{2}-f_{4})-\frac{1}{6}\rho v_{x}\quad\quad\text{for xlo;}
\end{eqnarray*}
\begin{eqnarray*}
&&\rho=\frac{1}{1+v_{y}}[f_{0}+f_{1}+f_{3}+2(f_{2}+f_{5}+f_{6})]\ ,\\
&&f_{4}=f_{2}-\frac{2}{3}\rho v_{y}\ ,\\
&&f_{7}=f_{5}+\frac{1}{2}(f_{1}-f_{3})-\frac{1}{6}\rho v_{y}\ ,\\
&&f_{8}=f_{6}-\frac{1}{2}(f_{1}-f_{3})-\frac{1}{6}\rho v_{y}\quad\quad\text{for yhi;}
\end{eqnarray*}
\begin{eqnarray*}
&&\rho=\frac{1}{1+v_{y}}[f_{0}+f_{1}+f_{3}+2(f_{4}+f_{7}+f_{8})]\ ,\\
&&f_{2}=f_{4}-\frac{2}{3}\rho v_{y}\ ,\\
&&f_{5}=f_{7}+\frac{1}{2}(f_{3}-f_{1})-\frac{1}{6}\rho v_{y}\ ,\\
&&f_{6}=f_{8}-\frac{1}{2}(f_{3}-f_{1})-\frac{1}{6}\rho v_{y}\quad\quad\text{for ylo.}
\end{eqnarray*}

In the style dirichlet, for the other walls we apply the rules:
\begin{eqnarray*}
&&v_{x}=1-\frac{f_{0}+f_{2}+f_{4}+2(f_{3}+f_{6}+f_{7})}{\rho}\ ,\\
&&f_{1}=f_{3}+\frac{2}{3}\rho v_{x}\ ,\\
&&f_{5}=f_{7}-\frac{1}{2}(f_{2}-f_{4})+\frac{1}{6}\rho v_{x}\ ,\\
&&f_{8}=f_{6}+\frac{1}{2}(f_{2}-f_{4})+\frac{1}{6}\rho v_{x}\quad\quad\text{for xlo;}
\end{eqnarray*}
\begin{eqnarray*}
&&v_{y}=\frac{f_{0}+f_{1}+f_{3}+2(f_{2}+f_{5}+f_{6})}{\rho}-1\ ,\\
&&f_{4}=f_{2}-\frac{2}{3}\rho v_{y}\ ,\\
&&f_{7}=f_{5}+\frac{1}{2}(f_{1}-f_{3})-\frac{1}{6}\rho v_{y}\ ,\\
&&f_{8}=f_{6}-\frac{1}{2}(f_{1}-f_{3})-\frac{1}{6}\rho v_{y}\quad\quad\text{for yhi;}
\end{eqnarray*}
\begin{eqnarray*}
&&v_{y}=1-\frac{f_{0}+f_{1}+f_{3}+2(f_{4}+f_{7}+f_{8})}{\rho}\ ,\\
&&f_{2}=f_{4}+\frac{2}{3}\rho v_{y}\ ,\\
&&f_{5}=f_{7}+\frac{1}{2}(f_{3}-f_{1})+\frac{1}{6}\rho v_{y}\ ,\\
&&f_{6}=f_{8}-\frac{1}{2}(f_{3}-f_{1})+\frac{1}{6}\rho v_{y}\quad\quad\text{for ylo.}
\end{eqnarray*}
\clearpage

\section{inlet/outlet openings for solute transport on a d2q4 lattice}\label{ap:inout_solute}

In the case of inlet openings, for the flux style the solute concentrations are updated by applying
the following rules \cite{book1}:
\begin{eqnarray*}
&&C=\frac{vC_{0}+DC_{-1}}{D+v}\quad\text{and}\quad g_{i}=g_{i}^{\mathrm{eq}}\ ,\quad i=1,\dots,4\ ,
\quad\quad\text{for xhi;}
\end{eqnarray*}
\begin{eqnarray*}
&&C=\frac{vC_{0}+DC_{-1}}{D+v}\quad\text{and}\quad g_{i}=g_{i}^{\mathrm{eq}}\ ,\quad i=1,\dots,4\ ,
\quad\quad\text{for yhi;}
\end{eqnarray*}
\begin{eqnarray*}
&&C=\frac{vC_{0}+DC_{1}}{D+v}\quad\text{and}\quad g_{i}=g_{i}^{\mathrm{eq}}\ ,\quad i=1,\dots,4\ ,
\quad\quad\text{for ylo.}
\end{eqnarray*}
For outlet openings the formulas become
\begin{eqnarray*}
&&C=\frac{vC_{0}-DC_{-1}}{-D+v}\quad\text{and}\quad g_{i}=g_{i}^{\mathrm{eq}}\ ,\quad i=1,\dots,4\ ,
\quad\quad\text{for xhi;}
\end{eqnarray*}
\begin{eqnarray*}
&&C=\frac{vC_{0}-DC_{1}}{-D+v}\quad\text{and}\quad g_{i}=g_{i}^{\mathrm{eq}}\ ,\quad i=1,\dots,4\ ,
\quad\quad\text{for xlo;}
\end{eqnarray*}
\begin{eqnarray*}
&&C=\frac{vC_{0}-DC_{-1}}{-D+v}\quad\text{and}\quad g_{i}=g_{i}^{\mathrm{eq}}\ ,\quad i=1,\dots,4\ ,
\quad\quad\text{for yhi;}
\end{eqnarray*}
\begin{eqnarray*}
&&C=\frac{vC_{0}-DC_{1}}{-D+v}\quad\text{and}\quad g_{i}=g_{i}^{\mathrm{eq}}\ ,\quad i=1,\dots,4\ ,
\quad\quad\text{for ylo.}
\end{eqnarray*}
$C_{0}$ is the concentration of the solution injected into the system; $v>0$ is the associated velocity. Given
a node belonging to the opening, $C_{1}$ and $C_{-1}$ represent the solute concentration for the nearest node 
inside the simulation domain.

For openings of type dirichlet, the solute concentration is updated using the following rules \cite{book1}:
\begin{eqnarray*}
&&C=\bar{C}\quad\text{and}\quad g_{1}=\bar{C}-C'\quad\quad\text{for xlo;}
\end{eqnarray*}
\begin{eqnarray*}
&&C=\bar{C}\quad\text{and}\quad g_{4}=\bar{C}-C'\quad\quad\text{for yhi;}
\end{eqnarray*}
\begin{eqnarray*}
&&C=\bar{C}\quad\text{and}\quad g_{2}=\bar{C}-C'\quad\quad\text{for ylo.}
\end{eqnarray*}
$\bar{C}$ is the prescribed solute concentration and $C'$ is the solute concentration obtained after the streaming 
step.
\clearpage

\section{Implementation of surface reaction on the lattice d2q4}\label{ap:d2q4}

\begin{figure*}[t]
\includegraphics[width=6cm]{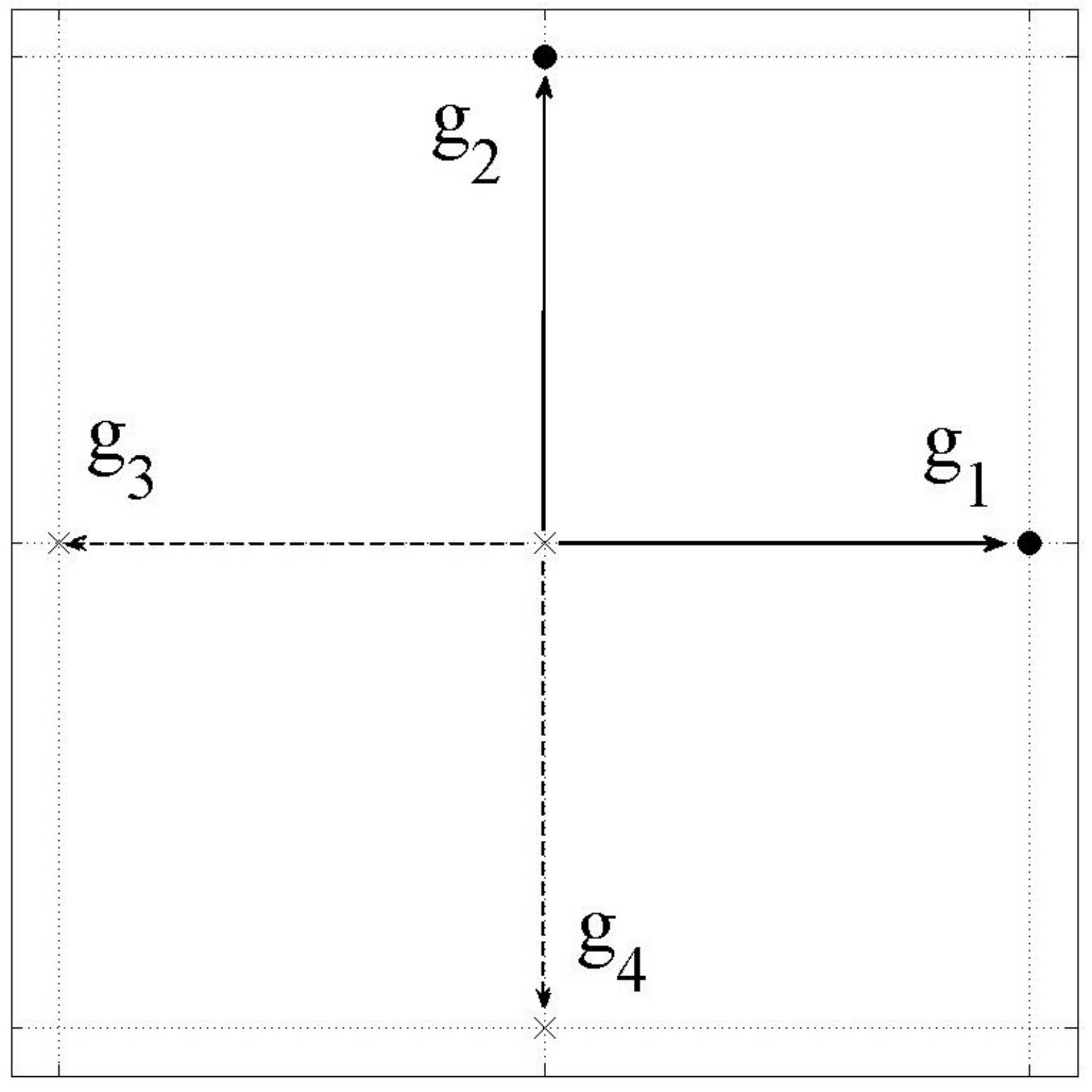}
\caption{\label{fig:vertex} 
Solute concentration modes at a solid corner on a d2q4 lattice; filled 
circles represent the fluid phase while crosses the solid boundary. Dashed
arrows are for outgoing modes from the fluid phase. After the streaming step,
both the incoming modes in the fluid phase indicated by solid arrows remain
undetermined.}
\end{figure*}

We provide details for the implementation of the surface reaction on the
lattice d2q4. The following rules are applied:

\begin{enumerate}
\item Two opposite unknown modes can not occur.

\item Two orthogonal unknown modes are possible at corners. The two $g_{i}$'s
are determined separately and the solute concentration is assumed to be
the average value $C=(C_{1}+C_{2})/2$. As an example, for the lower left
corner of Fig.~\ref{fig:vertex} we have
\begin{eqnarray*}
&&C_{1}=\frac{2g_{3}+k_{\mathrm{r}}C_{\mathrm{s}}}{k_{\mathrm{r}}+1/2}\quad\quad
g_{1}=-g_{3}+\frac{1}{2}C_{1}\ ,\\
&&C_{2}=\frac{2g_{4}+k_{\mathrm{r}}C_{\mathrm{s}}}{k_{\mathrm{r}}+1/2}\quad\quad
g_{2}=-g_{4}+\frac{1}{2}C_{2}\ ,\\
&&C=\frac{C_{1}+C_{2}}{2}\ .
\end{eqnarray*}

\item When the surface grows, the neighbors in the fluid phase of a solid node
have the same probability to become a solid node.

\item The deposited mass is subtracted from the solute concentration, as 
well as from the incoming modes in the fluid phase in equal parts \cite{snow}.

\item The cumulated mass of the solid node triggering surface growth is set to zero.
\end{enumerate}

\end{document}